\documentclass[reprint,
twocolumn,
amsmath,
amssymb,
aps,
physrev]{revtex4-2}
\usepackage{graphicx}
\usepackage{dcolumn}
\usepackage{bm}
\usepackage{lineno} 
\usepackage{enumitem} 
\usepackage{hyperref} 
\hypersetup{ 
	colorlinks=true,
	linkcolor=blue!80!black!,
	filecolor=magenta,      
	urlcolor=blue!50!black!,
	citecolor=red
}
\usepackage{labelschanged}
\usepackage{xcolor} 
\usepackage{amsfonts, amssymb, bbm}
\usepackage[bb=libus]{mathalpha}

\begin{document}
	\title{{Effects of super-Gaussian pulse shape and relative phase on dynamically assisted pair production in spatially inhomogeneous electric fields with frequency chirping}}
	
	\author{Abhinav Jangir}
	\email{2022rpy9087@mnit.ac.in}
	\author{Anees Ahmed}
	\email{anees.phy@mnit.ac.in}
	\affiliation{Department of Physics, Malaviya National Institute of Technology Jaipur, Jaipur, Rajasthan, India}
	\date{\today}
	
	\begin{abstract}
	We investigate the effects of super-Gaussian pulse shapes on dynamically assisted pair production in spatially inhomogeneous electric fields with frequency chirping within the framework of the (1+1)-dimensional Dirac–Heisenberg–Wigner formalism. The analysis is carried out for both single-color fields and dynamically assisted two-color combined fields by varying the chirp parameter, super-Gaussian pulse shape, and the spatial scale of the external field. Our results are presented in terms of the reduced momentum distribution and the total particle yield, revealing that the interplay between pulse shaping, spatial inhomogeneity, frequency chirping, and dynamical assistance can significantly modify the pair-production dynamics. In particular, chirping substantially enhances the absolute pair-production yield and reshapes the momentum spectrum, with chirping of the weak component being particularly effective, while simultaneous chirping of both components produces the largest yields. The super-Gaussian order has a relatively modest influence on strong-field production, while increasing the flat-top character of the pulse enhances spectral structures and momentum redistribution in the weak and dynamically assisted regimes, particularly under chirping. Finally, we show that the relative phase between the two field components provides an additional control parameter, with its influence becoming more pronounced for larger spatial scales leading to significant changes in the momentum-space structure. Overall, our results provide a useful reference for optimal control of dynamically assisted pair production in space- and time-dependent electric fields within a prescribed range of field parameters.
	\end{abstract}
	
	\maketitle
	\section{Introduction}\label{sec:Intro}
	The creation of electron-positron ($e^+e^-$) pairs from the vacuum in the presence of ultra-intense external electric fields is one of the most remarkable predictions of quantum electrodynamics (QED). This phenomenon, commonly known as the Schwinger effect, represents a fundamentally nonperturbative process in QED, where the vacuum becomes unstable against the spontaneous creation of charged particle pairs. The theoretical foundations of this effect trace back to Dirac's relativistic theory of the electron \cite{dirac1928quantum}, which not only established the framework of relativistic quantum mechanics but also predicted the existence of the positron. Subsequently, Sauter interpreted pair creation as a quantum tunneling process in the presence of a strong electric field \cite{sauter1931behavior}. Later, Heisenberg and Euler \cite{heisenberg1936folgerungen} calculated the leading pair production rate in weak electric field from the imaginary part of one-loop effective Lagrangian for spinor QED. In 1951, Schwinger derived the pair-production rate $\Gamma \sim \text{exp}(-\pi E_\text{cr}/E)$ for a constant electric field using the proper-time method \cite{schwinger1951gauge}, where $E_\text{cr} = m^2c^3/e\hbar \approx 1.3\times 10^{16}$ V/cm is the critical electric field strength ($m$ is electron mass and $-e$ is the electron charge) which corresponds to the laser intensity $I_\text{cr} \approx 4.3 \times 10^{29}$ W/cm$^{2}$. Subsequently, Nikishov \cite{nikishov1970pair} obtained the exact pair-production rate in a constant electric field by solving the Dirac equation using Green's function techniques, providing further insight into the nonperturbative nature of the process.
	
	Despite its profound theoretical importance, the Schwinger effect has not yet been observed experimentally due to the extremely large value of the critical field strength, which remains far beyond the capabilities of present-day laser facilities. Nevertheless, rapid advances in ultra-intense laser technologies continue to push the achievable field strengths closer to the nonperturbative regime. In particular, next-generation high-power laser facilities \cite{ELI, XFEL} are expected to reach intensities of the order of $I \approx 10^{26}$ W/cm$^{2}$ in the near future, significantly enhancing the prospects for the experimental observation of vacuum pair production. We refer the reader to Ref. \cite{Piazza2012extremely, fedotov2023advances} for recent investigations of high-energy processes in extremely intense laser fields within the frameworks of relativistic quantum dynamics, QED, and nuclear and particle physics.
	
	Over the past two decades, several methods, in addition to those discussed above, have been developed to investigate $e^+e^-$ pair production from the unstable vacuum in the presence of ultra-intense external electric fields. These include worldline instanton technique \cite{Gies2005pair, dunne2005worldline, Dunne2006worldline, schneider2016dynamically, dumlu2011complex}, the Wentzel-Kramers-Brillouin approach \cite{brezin1970pair, popov1972pair, oertel2019}, effective Lagrangian techniques \cite{dunne2005fields}, Furry-picture quantization \cite{aleksandrov2017momentum, aleksandrov2020pair}, quantum kinetic methods such as quantum Vlasov equation \cite{kluger1998quantum, aleksandrov2020kinetic, alkofer2001pair, nuriman2012enhanced, abdukerim2013effects, dumlu2011interference, dumlu2010schwinger, gong2020electron}, computational quantum field theory \cite{braun1999numerical, li2021study}, S-matrix theory \cite{muller2003differential, muller2003nonlinear, deneke2008bound}, and the Dirac-Heisenberg-Wigner (DHW) formalism \cite{blinne2014rotating, li2015effects, hebenstreit2010DHWvsQKE, hebenstreit2011particle}. 
	
	$e^+e^-$ pair production in spatially homogeneous but time-dependent electric fields constitutes a well-established area of study, with numerous field configurations investigated over the years, including Sauter pulses, multi-pulse structures, chirped pulses, and realistic laser envelopes. These investigations have revealed that the pair production rate and momentum distributions exhibit a pronounced dependence on the characteristics of the external field, such as the pulse duration, carrier-envelope phase, pulse separation, number of pulses, carrier frequency and frequency chirp, envelope shape, and field polarization \cite{nuriman2012enhanced, abdukerim2013effects, dumlu2010schwinger, dumlu2010stokes, dumlu2011interference, hebenstreit2009momentum, jangir2026carrier, olugh2019pair, chen2024asymmetric, olugh2020asymmetric, li2015effects, jangir2026dynamically}.                       

	To observe an appreciable pair-production rate via the Schwinger mechanism, electric field strengths comparable to the critical field are required \cite{gitman1996}. However, such extreme field intensities remain beyond the reach of current laser facilities. To circumvent this limitation, several catalytic mechanisms have been proposed to enhance vacuum pair production under subcritical field strengths \cite{bulanov2010multiple, piazza2009barrier, dunne2009catalysis, schutzhold2008dynamically, titov2012enhanced}. Among them, the dynamically assisted Schwinger effect (DASE) has emerged as one of the most effective approaches \cite{schutzhold2008dynamically, schneider2016dynamically, li2021enhanced, linder2015}. In this scheme, a strong low-frequency electric field is superimposed with a weak high-frequency field. The high-frequency component effectively lowers the tunneling barrier, resulting in an enhancement of the pair-production rate by several orders of magnitude compared with that of the strong field alone \cite{schutzhold2008dynamically}. Owing to its remarkable enhancement, the DASE has been extensively investigated in a variety of field configurations, with numerous studies examining the interplay between the two fields over a broad range of parameters in both temporally and spatially inhomogeneous backgrounds \cite{li2021enhanced, li2021study, ababekri2019effects, otto2015, nuriman2012enhanced, fey2012momentum, orthaber2011momentum, BraS2025, jangir2026dynamically, aleksandrov2018dase}.

	While studies of purely time-dependent electric fields have provided valuable insights into the nonperturbative dynamics of vacuum pair production, realistic laser fields are inherently inhomogeneous in both space and time. Incorporating spatial inhomogeneities is therefore essential not only for describing experimentally realizable field configurations in (3+1)-dimensions more accurately but also for developing a comprehensive theoretical understanding of vacuum pair production. However, such calculations are notoriously challenging, both analytically and computationally, owing to the increased complexity of the underlying dynamics. Fortunately, for electric fields with one-dimensional spatial dependence, the problem can be substantially simplified. Since pair production predominantly occurs along the direction of the electric field, the dynamics can, under appropriate conditions, be effectively reduced to (1+1)-dimensions by neglecting the particle momentum components transverse to the field direction \cite{hebenstreit2011particle, li2021study, ababekri2019effects, kohlfurst2018pondermotive}. This dimensional reduction provides a tractable framework for investigating the effects of spatial inhomogeneities while retaining the essential features of the pair-production dynamics. Several notable investigations of $e^+e^-$ pair production beyond the (1+1)-dimensional framework have also been reported in the literature; see, for example, Refs. \cite{kohlfurst2024pair, aleksandrov2020pair}.

	Spatial inhomogeneities introduce additional length scales that profoundly influence the pair-production process, giving rise to phenomena absent in spatially homogeneous backgrounds. In particular, spatial variations of the external field impart momentum to the Dirac sea, effectively increasing the energy gap between the negative- and positive-energy continua and thereby suppressing pair production. In contrast, temporal inhomogeneities supply energy to the vacuum, facilitating the creation of electron-positron pairs. As a result, the Schwinger effect is generally suppressed in spatially inhomogeneous electric fields compared with purely time-dependent field configurations \cite{dunne2005worldline, ilderton2014localisation, ruf2009pair, Gies2005pair, Gies2016critical}. Another important consequence of spatial inhomogeneities is the immediate acceleration experienced by $e^+e^-$ pairs upon their creation. In spatially localized electric fields, the momentum acquired by the particles depends on their creation position and the local field configuration, making the post-creation dynamics intrinsically sensitive to the spatial profile and extent of the field. As a result, spatial inhomogeneities influence not only the pair-production probability but also the subsequent evolution of the created particles, leaving characteristic signatures in the asymptotic momentum spectra \cite{osman2023efficient, bake2024enhanced, bake2025vacuum, Amat2023effect}. These spatial effects give rise to a range of nontrivial phenomena, including self-bunching of the produced particles \cite{hebenstreit2011particle}, where particles accumulate within narrow momentum intervals, and ponderomotive effects  \cite{kohlfurst2018pondermotive} induced by field gradients, which further reshape the final momentum distributions. 
	
	Different temporal field structures have been shown to significantly influence both the pair-production probability and the momentum distribution of the created particles in spatially inhomogeneous fields \cite{ababekri2019effects} . The role of spatial profiles has also attracted considerable attention. For example, Bake \textit{et al.} demonstrated that asymmetric spatial envelopes can give rise to nontrivial dynamical effects and, under suitable conditions, enhance the pair-production rate compared with conventional symmetric field configurations \cite{bake2024enhanced, bake2025vacuum, xu2026vacuum}. More recently, the influence of frequency chirping in spatially inhomogeneous electric fields has been explored in detail, revealing intricate temporal dynamics and multiple modulation mechanisms that can substantially modify the production process \cite{li2021enhanced, bake2025vacuum, ababekri2020chirp, mohamedsedik2021schwinger, osman2023efficient}. 
	Earlier studies in spatially homogeneous fields have established that the temporal pulse profile can strongly affect both the total production yield and the momentum spectra of the produced particles. In particular, several works have demonstrated that flat-top super-Gaussian pulse profiles provide a significant advantage over conventional Gaussian envelopes by sustaining the electric field near its peak value for a longer duration, thereby leading to enhanced pair production \cite{jangir2026carrier, abdukerim2013effects, zhou2024electron, otto2018assisted, xu2026effect}.
	
	In this work, we employ the DHW formalism in (1+1)-dimensions to systematically investigate frequency-chirped dynamically assisted $e^+e^-$ pair production in spatially inhomogeneous electric fields with super-Gaussian temporal profiles. While the effects of spatial inhomogeneity, pulse shaping, and frequency chirping have been investigated individually or in different combinations, their interplay in a dynamically assisted configuration with a non-Gaussian temporal envelope has received comparatively less attention. Our aim is to address this interplay within a unified framework and to identify how these field characteristics jointly control the pair-production yield and momentum distribution. 
	
	We begin by considering the individual strong and weak field components separately, allowing us to isolate the effects of chirp and spatial localization in each regime. For a range of chirp parameters and spatial scales, we analyze the corresponding reduced momentum distributions and reduced total yields, thereby establishing how the super-Gaussian pulse profile and spatial extent modify the production dynamics. We then consider the dynamically assisted two-color configuration, where the strong low-frequency and weak high-frequency components act simultaneously. A particular focus is placed on the location of the chirp: we systematically compare chirping the weak field alone, the strong field alone, and both components simultaneously. This comparison allows us to distinguish the qualitatively different roles played by frequency modulation of the two field components and to determine which configuration provides the most favorable conditions for pair production.
	
	In the two-color configuration, we further examine the dependence of pair production on the relative phase between the field components, providing insight into the coherent nature of the dynamical assistance mechanism in spatially inhomogeneous backgrounds. In this way, the present study goes beyond treating pulse shape, chirp, spatial localization, and phase dependence as separate effects and instead explores their combined influence on the nonperturbative production process. 
	
	The paper is organized as follows: In Sec. \ref{sec:Theoretical formalism}, we introduce the external electric-field model and briefly review the DHW formalism employed for our numerical analysis. In Sec. \ref{sec:Numerical_results_chirp_free_all_fields}, we present our numerical results for chirp-free field configurations. Section \ref{sec:Numerical_results_strong_chirp} is devoted to the effects of chirping in the strong field, while Sec. \ref{sec:Numerical_results_weak_chirp} examines the corresponding effects when chirping is applied to the weak field. In Sec. \ref{sec:Numerical_results_both_chirp}, we investigate the combined effect of chirping both the strong and weak field components simultaneously. The effects of a nonzero relative phase between the two field components are discussed in Sec. \ref{sec:non-zero_relative_phase}. Finally, we summarize our main findings and present our conclusions in Sec. \ref{sec:conclusion}.

	\section{Theoretical formalism}\label{sec:Theoretical formalism}
	In the present work, we study the $e^+e^-$ pair production in (1+1)-dimensions for spatially inhomogeneous, frequency-chirped electric fields with super-Gaussian temporal profiles. In particular, we focus on the combined influence of temporal pulse shaping, spatial inhomogeneity, and frequency chirping on pair production in one-color strong and weak fields, as well as in their dynamically assisted two-color configuration.
		\subsection{External electric field configuration} \label{subsec:field_Profile}
		Our model for the external electric field $E(x,t)$ is constructed as the sum of a strong, slowly varying component $E_{1s}(x,t)$ and a weak, rapidly varying component $E_{2w}(x,t)$, namely
		\begin{equation}\label{eq:electric_field} 
			\begin{split} 
				E(x,t) &= E_{1s}(x,t) + E_{2w}(x,t)\\ 
				&=E_{1s0}\,\mathcal{E}(x)\,\mathcal{F}(t)\,\text{cos}(b_1t^2 + \omega_1t)\\ 
				& \quad+E_{2w0}\,\mathcal{E}(x)\,\mathcal{F}(t)\,\text{cos}(b_2t^2 + \omega_2t + \varphi),
			\end{split} 
		\end{equation}
		where 
		\begin{equation} 
			\mathcal{E}(x) = \text{exp}\left(-\frac{x^2}{2\lambda^2}\right)\,\text{and}\,\mathcal{F}(t) = \text{exp}\left(-\frac{t^{2\nu}}{2\tau^{2\nu}}\right). 
		\end{equation}
		Here, $E_{1s0}$ and $E_{2w0}$ denote the peak amplitudes of the strong and weak field components, respectively. The parameter $\nu$ controls the order of the super-Gaussian temporal profile. The quantities $\omega_1$ and $\omega_2$ are the carrier frequencies of the two field components, while $\lambda$ and $\tau$ characterize the spatial and temporal extents of the external field. The coefficients $b_1$ and $b_2$ represent the frequency chirps of the strong and weak pulses, respectively. The parameter $\varphi$ is the relative phase between the strong and weak field components. Throughout most of this work, we consider the fields to be in phase by setting $\varphi =0$. The influence of a nonzero relative phase on the pair-production is investigated separately in Sec. \ref{sec:non-zero_relative_phase}.
		
		The external field is treated as an idealized standing-wave configuration formed by two counter-propagating coherent laser beams. In this picture, the magnetic-field contributions of the two waves cancel, leaving a purely spatially dependent electric field \cite{ababekri2020chirp, li2021enhanced}. Within this framework, particle production is assumed to occur predominantly along the field direction, and therefore the transverse momentum is set to $p_\perp = 0$. This effectively reduces the original (3+1)-dimensional problem to a (1+1)-dimensional description.
		
		Throughout this paper, natural units ($\hbar = c = 1$) are used. Furthermore, in our numerical calculations, a set of parameters characterizing the electric field \eqref{eq:electric_field} are fixed as
		\begin{equation}\label{eq:field_parameters}
			\begin{split}
				E_{1s0} &= 0.5E_\text{cr},\quad E_{2w0} = 0.15E_{1s0} = 0.075E_\text{cr},\\
				\omega_1 &= 0.1\,m,\quad \omega_2 = 7\omega_1 =  0.7\, m,\quad \tau = 25/m.
			\end{split}
		\end{equation}
		
		To characterize the dominant pair-production regime in the following analysis, we introduce the Keldysh adiabaticity parameter $\gamma = m\omega/eE$ \cite{keldysh1965ionization}, which distinguishes between the tunneling regime, $(\gamma \ll 1)$, and the multiphoton regime, $(\gamma \gg 1)$. For the field parameters specified in Eq. \eqref{eq:field_parameters}, the chirp-free strong-field component is characterized by $\gamma_s = m\omega_1/eE_{1s0} = 0.2$, placing it firmly in the tunneling regime. In contrast, the chirp-free weak-field component has $\gamma_w = m\omega_2/eE_{2w0} = 9.34$ corresponding to the multiphoton regime. Thus, the two field components operate in distinct pair-production regimes, providing the characteristic separation between the tunneling and multiphoton scales required for dynamically assisted pair production.
		
		For a nonzero chirp, the instantaneous frequency of the field becomes time dependent and is given by $\omega_\text{eff} = \omega + bt$, where $b$ denotes the linear chirp parameter. Accordingly, the Keldysh parameter also acquires an explicit time dependence through the instantaneous frequency. In the present analysis, we restrict the chirp to the \textit{normal-chirp} regime \cite{olugh2019pair, li2021enhanced}, for which the chirp parameter is parameterized as $b_i = a_i\omega_i/\tau$ with $0\leq a_i\leq1$ ($i=1,2$). The upper bound $a_i=1$ corresponds to $b_1^\text{max}=\omega_1/\tau = 0.004\,m^2$ for $E_{1s}$, and $b_2^\text{max}=\omega_2/\tau = 0.028\,m^2$ for $E_{2w}$, respectively. These values correspond to relatively strong chirping and are deliberately included in our analysis to explore the qualitative trends and dynamical modifications that emerge in the strongly chirped regime.
	
		\subsection{DHW formalism}\label{subsec:DHW_formalism}
		The present work is based on the DHW formalism, which is a quantum kinetic theory formulated in terms of the Wigner function to characterize relativistic phase-space distributions \cite{vasak1987quantum, birula1991phase}. The formalism has been widely adopted in studies of vacuum pair production in arbitrary electromagnetic fields \cite{hebenstreit2011particle, kohlfurst2018pondermotive, ababekri2019effects, li2021enhanced, bake2025vacuum, bake2024enhanced, ababekri2020chirp, mohamedsedik2021schwinger, osman2023efficient, diez2023identifying, mohamedsedik2023phase}. Detailed derivations and systematic developments of the DHW formalism can be found in Refs. \cite{HebenstreitPhD, KohlfurstPhD}. Since a comprehensive review of the formalism is beyond the scope of the present work, we restrict ourselves here to outlining only the essential aspects of the method. 
		
		We start from the gauge-invariant density operator of the system,
		\begin{equation}\label{eq:gauge-invariant density operator}
			\hat{\mathcal{C}}_{\alpha\beta}(r,s) = \mathcal{U}(A,r,s)[\bar{\psi}_\beta(r - s/2),\psi_\alpha(r + s/2)],
		\end{equation}
		where $\psi_\alpha(x)$ is spinor-valued Dirac field of the electron, $r$ the center of mass, and $s$ is the effective coordinate. To ensure the gauge invariance of the density operator, we introduce Wilson line factor
		\begin{equation}\label{eq:Wilson line factor}
			\mathcal{U}(A,r,s) = \text{exp}\left(ies\int^{1/2}_{-1/2}d\xi A(r + \xi s)\right)
		\end{equation}
		which depends on the elementary charge $e$ and background gauge field $A$. With the Hartree approximation of the background gauge field i.e.,
		\begin{equation}\label{eq:Hartree approximation}
			F^{\mu\nu}(x) \approx \langle\hat{F}^{\mu\nu}(x)\rangle,
		\end{equation}
		no path ordering is required, and in a given Lorentz frame and gauge, the background gauge field A(\textbf{x},t) is a fixed $C$-number valued function rather than a $Q$-number operator \cite{vasak1987quantum, KohlfurstPhD}. The covariant Wigner operator is defined as the Fourier transform of the density operator in Eq. \eqref{eq:gauge-invariant density operator} from the $s$-space to the $p$-space, yielding
		\begin{equation}\label{eq:covariant Wigner operator}
			\hat{\mathcal{W}}_{\alpha\beta}(r,p) = \frac{1}{2}\int d^4s\,\,\text{e}^{ips}\hat{\mathcal{C}}_{\alpha\beta}(r,s).
		\end{equation}
		This operator incorporates the quantum fluctuations of the electron field, while the laser field is treated classically. 
		
		On taking the vacuum expectation value of Eq. \eqref{eq:covariant Wigner operator}, one obtains the covariant Wigner function
		\begin{equation}\label{eq:covariant Wigner function}
			\mathbb{W}(r,p) = \langle\Phi|\hat{\mathcal{W}}(r,p)|\Phi\rangle,
		\end{equation}
		where $|\Phi\rangle $ and $\langle\Phi|$ denote the initial and final vacuum state. 
		
		The mean-field approximation considerably simplifies the computation of the vacuum expectation value of the Wigner operator, as the electromagnetic field tensor factored out in the correlation function,
		\begin{equation}
			\langle\Phi|F_{\mu\nu}\hat{\mathcal{C}}|\Phi\rangle = 	F_{\mu\nu}\langle\Phi|\hat{\mathcal{C}}|\Phi\rangle
		\end{equation}
		Within the Dirac algebra, the Wigner function can be expanded in terms of 16 covariant Wigner coefficients, satisfying
		\begin{equation}
		\mathbb{W} = \frac{1}{4}(\mathbbm{1}\mathbb{S} + i\gamma_5\mathbb{P} + \gamma^\mu\mathbb{V}_\mu + \gamma^\mu\gamma_5\mathbb{A}_\mu + \sigma^{\mu\nu}\mathbb{T}_{\mu\nu}).
		\end{equation}
		Here, $\mathbb{S}$, $\mathbb{P}$, $\mathbb{V}_{\mu}$, $\mathbb{A}_{\mu}$, and $\mathbb{T}_{\mu\nu}$ represent the scalar, pseudoscalar, vector, axial-vector, and tensor components, respectively. In particular, $\mathbb{S}$ corresponds to the mass density, $\mathbb{P}$ to the condensate density, $\mathbb{V}_{\mu}$ to the fermionic current density, $\mathbb{A}_{\mu}$ to the polarization density, and $\mathbb{T}_{\mu\nu}$ to the electric dipole moment density \cite{birula1991phase}. The equal-time Wigner function is defined thorough the energy average of the covariant Wigner function, 
		\begin{equation}
			\mathbb{w}(\textbf{x},\textbf{p},t) = \int \frac{dp_0}{2\pi}\mathbb{W}(r,p)
		\end{equation}
		where $\textbf{x}$, $\textbf{p}$ denote the particle position and kinetic momentum, respectively. In the context of vacuum pair production, the dynamical evolution of the system is completely characterized by the equal-time Wigner function. Consequently, all physical quantities considered in this work are expressed in terms of its components.
		
		We now apply the DHW formalism in a (1+1)-dimensional setting, consisting of one spatial coordinate and a time-dependent electric field $E(x,t)$ specified by Eq. \eqref{eq:electric_field}. Under this dimensional reduction, only  $\mathbb{S}$, $\mathbb{P}$, and $\mathbb{V}_{\mu}$ $(\mu=0,1)$ remain nonvanishing. In the equal-time formulation, the corresponding physical quantities are denoted by lowercase letters, leaving four independent Wigner components: $\mathbb{s}$, $\mathbb{v}_0$, $\mathbb{v}_x$, and $\mathbb{p}$. Consequently, the full system of equations of motion simplifies to a coupled set of four equations \cite{birula1991phase, KohlfurstPhD},
		\begin{equation}\label{eq:DHW_equations}
			\begin{split}
			D_t\mathbb{s} - 2p_x\mathbb{p} &= 0,\\[6pt]
			D_t\mathbb{v}_0 + \partial_x\mathbb{v}_x &= 0,\\[6pt]
			D_t\mathbb{v}_x + \partial_x\mathbb{v}_0 &= -2m\mathbb{p},\\[6pt]
			D_t\mathbb{p} + 2p_x\mathbb{s} &= 2m\mathbb{v}_x,
			\end{split}
		\end{equation}
		where $D_t$ denotes a differential operator composed of the ordinary time derivative together with a pseudo-differential operator, given by
		\begin{equation}
				D_t  = \partial_t + e\int^{1/2}_{-1/2} d\xi E_x (x + i\xi \partial_{p_{x}}, t)\partial_{p_x}.
		\end{equation}
		The corresponding vacuum initial conditions \cite{KohlfurstPhD} are given by 
		\begin{equation}
			\bigg[\mathbb{s},\mathbb{v}_0,\mathbb{v}_x,\mathbb{p}\bigg]_\text{vac}  = \bigg[-\frac{2m}{\Omega}, 0,-\frac{2p_x}{\Omega}, 0\bigg]
		\end{equation}
		where $\Omega = \sqrt{p_x^2 + m^2}$ is the energy of a particle. On subtracting the vacuum contributions, one obtains the modified Wigner components as
		\begin{equation}
			\tilde{\mathbb{w}} = \mathbb{w} - \mathbb{w}_\text{vac}.
		\end{equation}
		The Wigner components are not observable physical quantities. 
		\begin{figure*}[tbh]
			\centering
			\includegraphics[width=\textwidth]{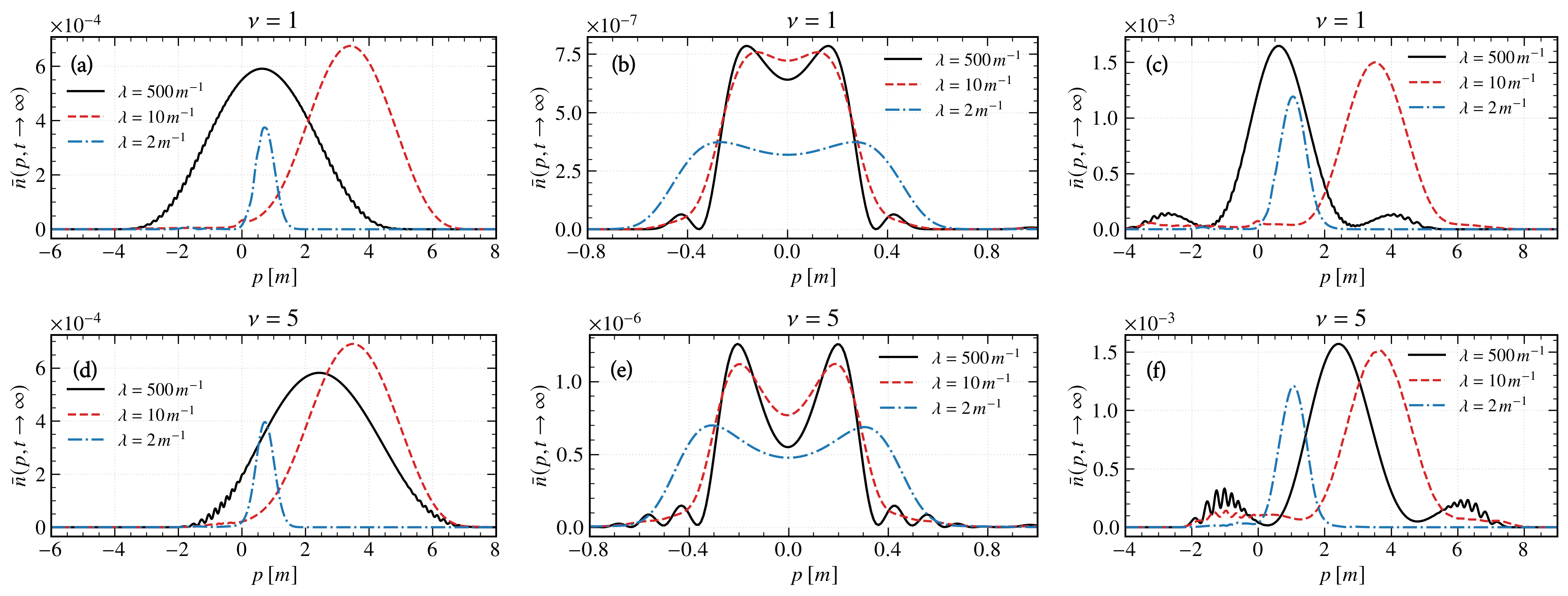}
			\caption{Reduced momentum distribution for different spatial scales. The first column [panels (a),(d)], second column [panels (b),(e)], and third column [panels (c),(f)] correspond to the one-color strong field $E_{1s}(x,t)$, one-color weak field $E_{2w}(x,t)$, and two-color combined field $E(x,t)$, respectively. The first row [panels (a), (b), and (c)] and second row [panels (d), (e), and (f)] represent the cases $\nu=1$ and $\nu=5$, respectively. Chirping is not included in any case i.e., $b_1 = b_2 = 0$. The other field parameters are given in Eq.~\eqref{eq:field_parameters}.}
			\label{fig:np_chirp_free_SWC_fields}
		\end{figure*}
		The particle number density in phase space is formally defined as \cite{hebenstreit2011particle} 
		\begin{equation}\label{eq:n(x,p,t) number density in phase space}
			n(x,p_x,t) = \frac{m\tilde{\mathbb{s}}(x,p_x,t) + p_x \tilde{\mathbb{v}}_x(x,p_x,t)}{\Omega(p_x)}.
		\end{equation}
		Further, the position and momentum distribution of the created particles can be derived from $n(x,p_x,t)$ as follows \cite{KohlfurstPhD, ababekri2019effects, ababekri2020chirp}:
		\begin{equation}\label{eq:spatial distribution}
			n(x,t) = \int dp_x n(x,p_x,t),
		\end{equation}
		\begin{equation}\label{eq:momentum distribution}
			n(p_x,t) = \int \frac{dx}{2\pi} n(x,p_x,t).
		\end{equation}
		The total particle yield is obtained by integrating over the entire phase space,
		\begin{equation}\label{eq:total particle yield}
			N(t) = \int \frac{dx}{2\pi}\int dp_x n(x,p_x,t).
		\end{equation}
		In the following sections, the numerical results are presented in terms of the reduced quantities at $t\rightarrow\infty$: $\bar{n}(p_x,t\rightarrow\infty) \equiv n(p_x,t\rightarrow\infty)/\lambda$ and $\bar{N}(t\rightarrow\infty) \equiv N(t\rightarrow\infty)/\lambda$	which are introduced to incorporate the nontrivial dependence on the spatial scale $\lambda$. Furthermore, for notational simplicity, the subscript in $p_x$ will be omitted hereafter, and the momentum will be denoted simply by $p$.
		
		\subsection{Numerical implementation}
		We solve the coupled equations \eqref{eq:DHW_equations} using a Fourier pseudo-spectral method based on fast Fourier transforms (FFTs) \cite{KohlfurstPhD}.
		The system of ordinary differential equations is integrated over the time interval $t\in[-\eta\tau,\eta\tau]$, where $\eta>0$ is chosen sufficiently large, using the variable-order Adams--Bashforth--Moulton predictor--corrector algorithm implemented in MATLAB's \texttt{ode113} solver with $\mathrm{RelTol}=10^{-10}$ and $\mathrm{AbsTol}=10^{-10}$.
		
		The numerical domain and grid resolution were tested for convergence by varying the momentum cutoff $L_p$, the momentum resolution $N_p$, spatial domain size $L_x$, and the spatial resolution $N_x$.  These numerical parameters were chosen such that further enlargement of the computational domain or refinement of the numerical grid produces no appreciable change in the calculated momentum spectra or total particle yields. Furthermore, the momentum domain was selected sufficiently large to suppress spurious oscillations and Gibbs-type artifacts associated with the Fourier representation. The numerical parameters adopted in the calculations are consistent with those used in Ref.~\cite{ababekri2019effects} and are summarized in Table \ref{tab:numerical_parameters}.
		
		\begin{table}[tbh]
			\caption{Numerical parameters used in our calculations.}
			\label{tab:numerical_parameters}
			\begin{ruledtabular}
				\begin{tabular}{ccccc}
					$\lambda[\lambda_c]$ & $N_x$ & $N_p$ & $L_x[\lambda_c]$ & $L_p[m]$ \\
					\hline
					$\lambda \leq 10$ & 512 & 2048 & $150+3.5\lambda$ & 40 \\
					$10 < \lambda < 100$ & 256 & 2048 & $150+3.5\lambda$ & 40 \\
					$\lambda \geq 100$ & 128 & 2048 & $150+3.5\lambda$ & 40 \\
				\end{tabular}
			\end{ruledtabular}
		\end{table}

	\section{Numerical results: Chirp-free fields}\label{sec:Numerical_results_chirp_free_all_fields}
	
	In this section, we present the results for the one-color field configurations, $E_{1s}(x,t)$ and $E_{2w}(x,t)$, as well as for the corresponding two-color combined field $E(x,t)$. For clarity, we first consider the unchirped case, with the effects of frequency chirping discussed separately in the subsequent sections. Fig. \ref{fig:np_chirp_free_SWC_fields} presents the reduced momentum distribution, $\bar{n}(p,t\rightarrow\infty)$, for the different field configurations, illustrating its dependence on $\lambda$ and $\nu$. Three spatial scales, $\lambda=500~m^{-1}$, $10~m^{-1}$, and $2~m^{-1}$, are considered throughout the analysis.
	\begin{figure*}[tbh]
		\centering
		\includegraphics[width=\textwidth]{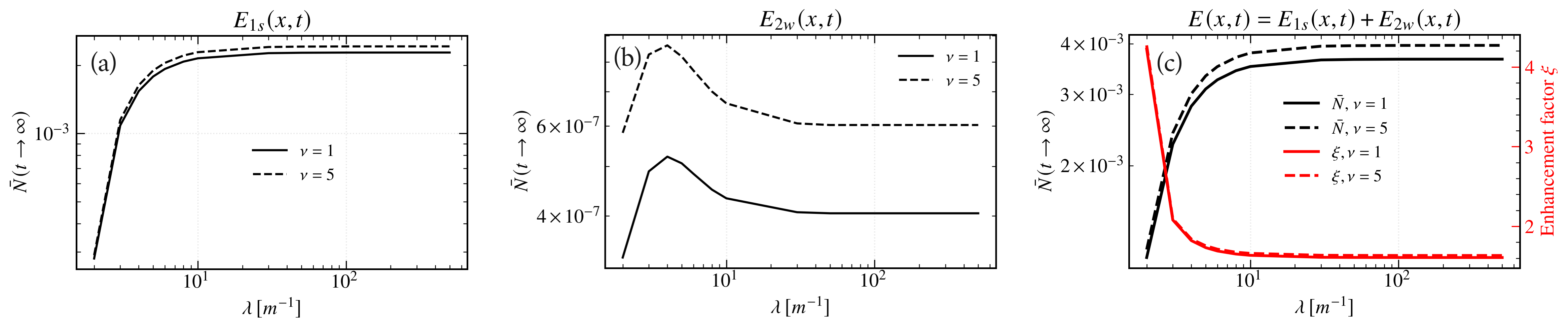}
		\caption{Reduced total yield for one-color strong field $E_{1s}(x,t)$ (panel (a)), one-color weak field $E_{2w}(x,t)$ (panel (b)), and two-color combined field $E(x,t)$ (panel (c)) as a function of spatial scales. Panel (c) also shows the enhancement factor on the secondary (right) axis. Results are presented for $\nu=1$ (solid lines) and $\nu=5$ (dashed lines). Chirping is not included in any case i.e., $b_1 = b_2 = 0$. The other field parameters are given in Eq.~\eqref{eq:field_parameters}.}
		\label{fig:N_chirp_free_SWC_fields}
	\end{figure*}
	
	For the one-color strong field, shown in Figs. \ref{fig:np_chirp_free_SWC_fields}(a) and \ref{fig:np_chirp_free_SWC_fields}(d), the momentum distribution exhibits a pronounced dependence on the spatial scale. This configuration, with the corresponding parameter set defined in Eq. \eqref{eq:field_parameters}, was studied in detail in Ref. \cite{ababekri2020chirp} for Gaussian temporal envelope. We therefore do not repeat that analysis in detail, but briefly discuss the momentum distributions to establish a baseline for the subsequent results for variation in $\lambda$ and $\nu$. For the nearly spatially homogeneous case, $\lambda=500\ m^{-1}$, the distribution is relatively broad and possesses a single dominant maximum. Reducing the spatial scale to $\lambda=10\ m^{-1}$ shifts the dominant structure towards larger positive momenta and modifies its width. For the strongest spatial inhomogeneity, $\lambda=2\ m^{-1}$, the broad distribution is replaced by a much narrower and more localized peak due to the extremely small spatial extent of the field, which limits the distance over which the electric field can perform work on the produced particles and, consequently, restricts the momentum gain. A comparison between Figs. \ref{fig:np_chirp_free_SWC_fields}(a) and \ref{fig:np_chirp_free_SWC_fields}(d) further shows that increasing $\nu$ from $1$ to $5$ changes the momentum profile even in the absence of chirp. In particular, for $\lambda=500\ m^{-1}$, the distribution moves towards larger positive momentum and the distribution becomes somewhat broader for $\nu=5$. This behavior arises from the flatter super-Gaussian envelope, which sustains pair production over a longer time and modifies the accumulated particle momentum. Its sharper temporal switching also enhances high-frequency components, allowing pairs to populate a wider momentum range \cite{aleksandrov2025switchoff, xu2026effect}. The flat-top envelope effects for $\lambda=10\,m^{-1}$ and $2\,m^{-1}$ remain qualitatively similar, as the strong spatial inhomogeneity remains the dominant mechanism shaping the spectrum, although their detailed peak positions, widths, and magnitudes are slightly modified.

	The behavior of the one-color weak field, shown in Figs. \ref{fig:np_chirp_free_SWC_fields}(b) and \ref{fig:np_chirp_free_SWC_fields}(e), is qualitatively different from that of the strong field. Here, the distributions for the two relatively large spatial scales, $\lambda=500\,m^{-1}$ and $10\,m^{-1}$, exhibit a characteristic two-lobed structure centered around positive and negative momenta. In contrast to the slowly-oscillating strong-field case, the rapidly-oscillating weak-field distributions remain approximately symmetric about $p=0$. This can be understood from the fact that the weak field has a much higher carrier frequency and therefore contains many more oscillations within the pulse duration. Consequently, the positive and negative contributions to the temporal integral of the electric field largely cancel, resulting in a negligible asymptotic vector potential, $A(\infty)\simeq0$. The asymptotic kinetic momentum, $p_\infty=k-eA(\infty)$, therefore remains approximately identical to the canonical momentum, and the momentum distribution is not appreciably displaced from $p=0$. In contrast, the lower-frequency strong field contains fewer carrier oscillations within the same envelope, leading to an incomplete cancellation and a finite $A(\infty)$, which produces the observed momentum displacement. 
	
	As the spatial scale is reduced to $\lambda=2\,m^{-1}$, the two-lobed structure becomes broader and its overall magnitude is substantially reduced. The peak splitting observed in the large spatial scale, $\lambda=500\,m^{-1}$, can be attributed to interference effects between particles produced at different temporal regions of the field \cite{dumlu2011interference}. In contrast, for the strongly spatially inhomogeneous cases, $\lambda=10\,m^{-1}$ and $2\,m^{-1}$, the spatial field gradient generates a ponderomotive force that drives produced particles away from the strongest-field region, thereby broadening and splitting the final momentum spectrum \cite{kohlfurst2018pondermotive, li2021enhanced}. This effect becomes more pronounced with increasing super-Gaussian order $\nu$, consistent with the stronger ponderomotive force associated with higher-order super-Gaussian field profiles, as recently reported in Ref. \cite{xu2026effect}.
	
	The influence of $\nu$ on the weak-field distribution can be seen by comparing Figs. \ref{fig:np_chirp_free_SWC_fields}(b) and \ref{fig:np_chirp_free_SWC_fields}(e). For $\nu=5$, the two-lobed structure remains clearly visible for $\lambda=500\ m^{-1}$ and $10\ m^{-1}$, although its detailed shape and amplitude are modified. The distribution for $\lambda=2\ m^{-1}$ again shows a broader and weaker profile. Importantly, the qualitative symmetry of the weak-field momentum distribution is preserved when $\nu$ is increased, in contrast to the pronounced momentum displacement observed for the strong field. 
	\begin{figure*}[tbh]
		\centering
		\includegraphics[width=\textwidth]{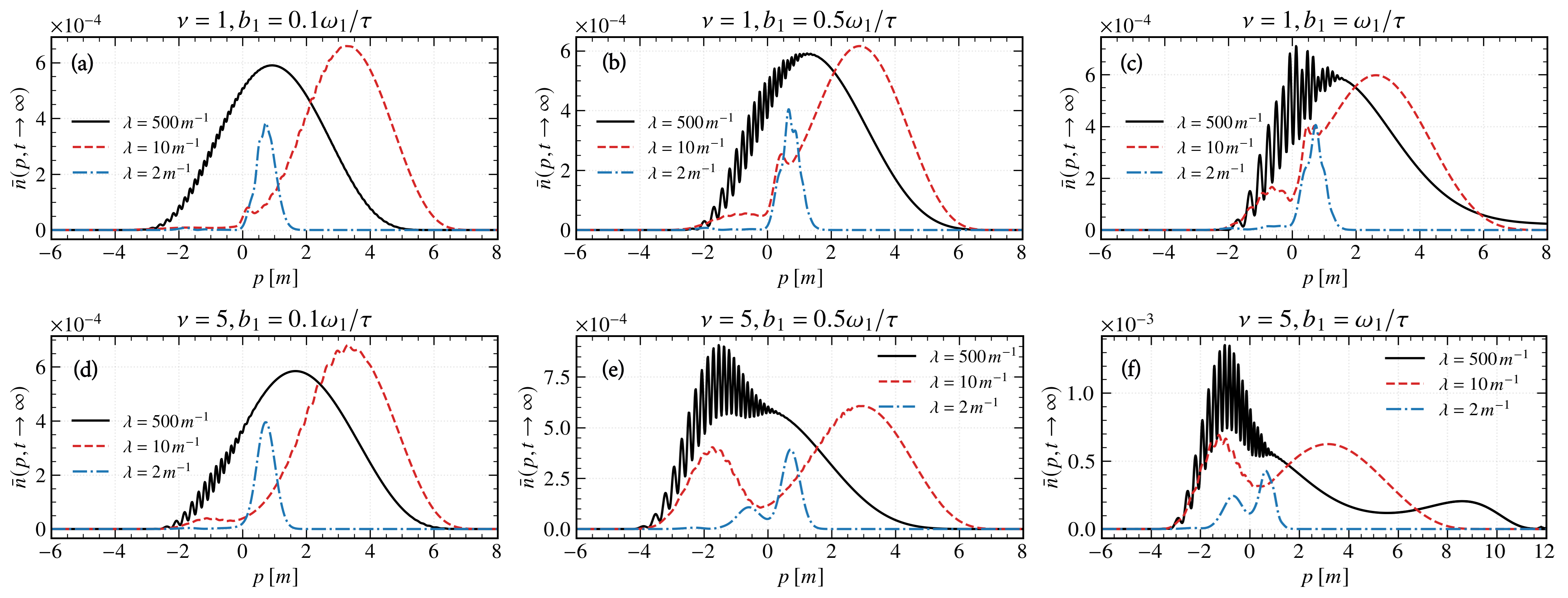}
		\caption{Reduced momentum distribution for different spatial scales in one-color strong field $E_{1s}(x,t)$. The first column [panels (a),(d)], second column [panels (b),(e)], and third column [panels (c),(f)] correspond to the chirp $b_1[\omega_1/\tau] = 0.1,0.5$ and $1$, respectively. The first row [panels (a), (b), and (c)] and second row [panels (d), (e), and (f)] represent the cases $\nu=1$ and $\nu=5$, respectively. The other field parameters are given in Eq.~\eqref{eq:field_parameters}.}
		\label{fig:np_strong_chirp_strong_field}
	\end{figure*}
	
	The effect of combining the two field components is illustrated in Figs. \ref{fig:np_chirp_free_SWC_fields}(c) and \ref{fig:np_chirp_free_SWC_fields}(f). The two-color field produces momentum distribution that are substantially different from either of the corresponding one-color cases. For $\lambda=500\ m^{-1}$, the combined field exhibits a pronounced peak at positive momentum, accompanied by oscillatory tails extending on both sides of the main peak. As the spatial scale is reduced to $\lambda=10\ m^{-1}$, the dominant peak is shifted towards larger positive momentum and its magnitude changes considerably. For $\lambda=2\ m^{-1}$, the distribution develops a relatively narrow peak at intermediate positive momentum. The oscillatory tails gradually disappear as the spatial scale decreases from $\lambda = 500\ m^{-1}$ to $2\ m^{-1}$. These features demonstrate that the weak, rapidly varying field does not simply provide an additive contribution to the momentum spectrum. Instead, its presence modifies the momentum distribution generated by the strong field, leading to a nontrivial redistribution of the produced pairs. A similar behavior is observed for $\nu=5$ in Fig. \ref{fig:np_chirp_free_SWC_fields}(f). The distribution retain the characteristic dependence on $\lambda$, but the structures become more pronounced and additional oscillations appear, particularly for the larger $\lambda$. The comparison of Figs. \ref{fig:np_chirp_free_SWC_fields}(c) and \ref{fig:np_chirp_free_SWC_fields}(f) therefore shows that increasing $\nu$ can significantly modifies the fine structure of the spectrum, while the overall dependence on $\lambda$ remains robust. 

	To further analyze our study, we investigate the corresponding reduced total yield $\bar{N}(t\rightarrow\infty)$ in presence of the one-color strong $E_{1s}(x,t)$, weak $E_{2w}(x,t)$, and two-color combined field $E(x,t) = E_{1s}(x,t) + E_{2w}(x,t)$, all in chirp free cases. The corresponding reduced total yields as a function of $\lambda$ with variation in $\nu$ for different field configurations, is shown in Fig. \ref{fig:N_chirp_free_SWC_fields}. The reduced total yield provides a complementary measure to the momentum distributions discussed above and allows the overall effect of $\lambda$ to be quantified. For the one-color strong field, Fig. \ref{fig:N_chirp_free_SWC_fields}(a), the yield increases rapidly with increasing $\lambda$ and approaches a nearly saturated value at large spatial scales. The results for $\nu=1$ and $\nu=5$ remain close to each other over the entire range, with the $\nu=5$ case giving a slightly larger yield at large $\lambda$. 

	For the weak field, Fig. \ref{fig:N_chirp_free_SWC_fields}(b), the dependence on $\lambda$ is qualitatively different. The yield initially increases as $\lambda$ is increased, reaches a maximum at an intermediate value of $\lambda$, and subsequently decreases towards a nearly constant value for larger $\lambda$. The yield for $\nu=5$ remains consistently larger than that for $\nu=1$, indicating a stronger sensitivity of the weak-field contribution to $\nu$.

	The combined-field results in Fig. \ref{fig:N_chirp_free_SWC_fields}(c) exhibit a pronounced increase in the reduced total yield with increasing spatial scale $\lambda$, followed by saturation in the weakly inhomogeneous regime. The yield obtained from the combined field is substantially larger than those produced by the individual one-color fields, indicating a significant enhancement arising from the interplay between the strong and weak field components.
	
	To quantify the efficiency of the dynamically assisted two-color configuration $E(x,t)$, we consider the enhancement factor, shown on the secondary axis of Fig. \ref{fig:N_chirp_free_SWC_fields}(c). It is defined as \cite{aleksandrov2018dase, li2021study} $\xi = {\bar{N}_{1s + 2w}}/({\bar{N}_{1s} + \bar{N}_{2w}})$,
	where, $\bar{N}_{1s+2w}$ denotes the reduced total yield obtained from the combined field, while $\bar{N}_{1s}$ and $\bar{N}_{2w}$ correspond to the yields obtained from the strong and weak fields separately. Consequently, $\xi$ provides a measure of the enhancement associated with the combined field relative to the individual contributions of its strong and weak components. The enhancement factor is found to be largest for the smallest spatial scales, indicating that the dynamically assisted enhancement becomes particularly pronounced in the presence of strong spatial inhomogeneity. As $\lambda$ increases, the enhancement factor decreases and eventually approaches an approximately constant value in the weakly inhomogeneous limit. A slightly larger enhancement is observed for $\nu=5$ compared with $\nu=1$ over the considered range of $\lambda$.

	\section{Numerical results: Chirp applied to $E_{1s}(x,t)$ only}\label{sec:Numerical_results_strong_chirp}
	In this section, we present the results for the one-color strong field $E_{1s}(x,t)$ and the two-color combined field $E(x,t)$. In both the cases, we consider the chirping only for $E_{1s}(x,t)$, i.e., $b_1 = 0.1\omega_1/\tau, 0.5\omega_1/\tau$ and $\omega_1/\tau$, and $b_2=0$.
	\subsection{One-color strong field $E_{1s}(x,t)$} \label{subsec:strong_chirp_strong_field}

\begin{figure}[tbh]
		\centering
		\includegraphics[width=0.9\linewidth]{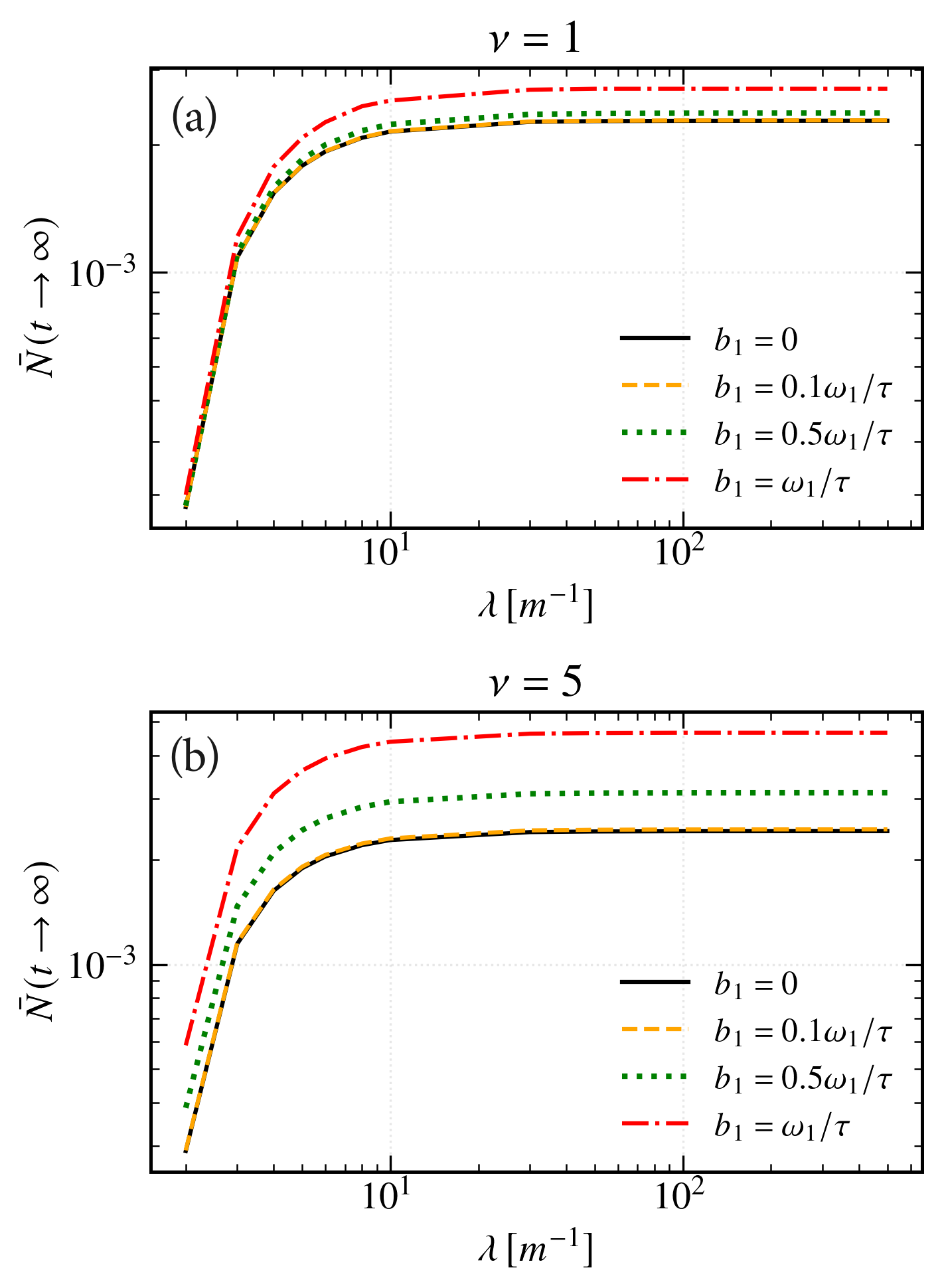}
		\caption{Reduced total yield for one-color strong field $E_{1s}(x,t)$ with different chirp as a function of spatial scales. Results are presented for $\nu=1$ (panel (a)) and $\nu=5$ (panel (b)). The other field parameters are given in Eq.~\eqref{eq:field_parameters}.}
		\label{fig:N_strong_field}
	\end{figure}

	In this subsection, the results are discussed for chirped one-color strong field with $\lambda$ and $\nu$ varied. Fig. \ref{fig:np_strong_chirp_strong_field} shows the corresponding reduced momentum distributions for three nonzero chirp parameters, $b_1 = 0.1\omega_1/\tau, 0.5\omega_1/\tau$ and $\omega_1/\tau$, while Fig. \ref{fig:N_strong_field} presents the corresponding reduced total yield.
	
	For $\nu=1$, the chirp-free distribution considered previously consists primarily of a broad dominant peak for $\lambda=500\,m^{-1}$ [Fig. \ref{fig:np_chirp_free_SWC_fields}(a)], which becomes progressively narrower and shifts towards smaller positive momenta as $\lambda$ is reduced. Introducing a finite chirp substantially modifies this structure. For $\lambda=500\,m^{-1}$, the distribution remains dominated by a single broad peak, but oscillatory structures develop on its negative-momentum side as the chirp is increased. These oscillations become particularly pronounced for $b_1 = \omega_1/\tau$, where the distribution develops rapid large oscillations in $p\sim (-2m,2m)$. The observed oscillatory structures in the momentum distribution can be understood as arising from interference between particle produced at the oppositely signed, dominant peaks of the temporal electric field \cite{dumlu2010schwinger, dumlu2011interference}. For $\lambda=10\,m^{-1}$, the main peak remains centered at positive momentum, with its position and width being modified by the chirp, while additional oscillations emerge on its left-momentum side. In contrast, for the strongly inhomogeneous case $\lambda = 2\,m^{-1}$, the distribution remains comparatively narrow and localized, although its peak position and magnitude are slightly altered as the chirp increases. Thus, increasing the chirp primarily introduces increasingly pronounced fine structures for the larger spatial scales, whereas strong spatial inhomogeneity tends to maintain a more localized momentum distribution.
	\begin{figure*}[tbh]
		\centering
		\includegraphics[width=\textwidth]{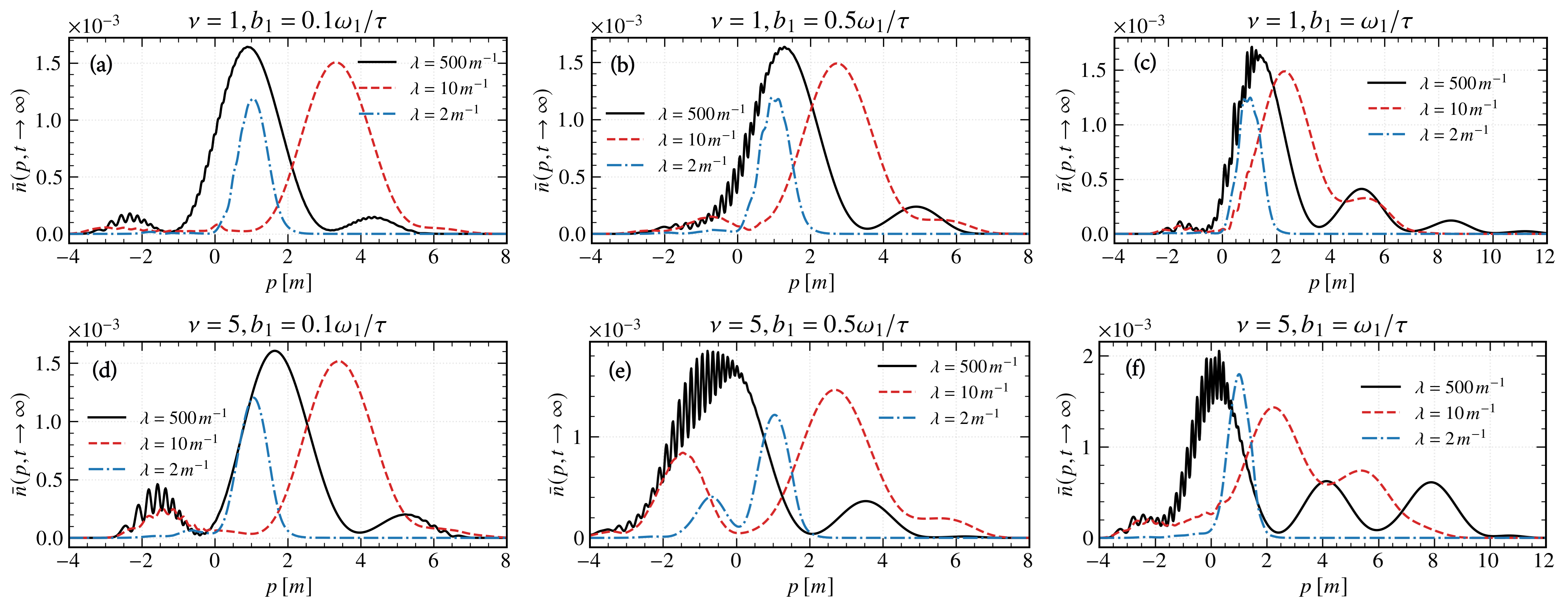}
		\caption{Reduced momentum distribution for different spatial scales in two-color combined field $E(x,t)$ with chirp applied to $E_{1s}(x,t)$ only. The first column [panels (a),(d)], second column [panels (b),(e)], and third column [panels (c),(f)] correspond to the chirp $b_1[\omega_1/\tau] = 0.1,0.5$ and $1$, respectively. The first row [panels (a), (b), and (c)] and second row [panels (d), (e), and (f)] represent the cases $\nu=1$ and $\nu=5$, respectively. The other field parameters are given in Eq.~\eqref{eq:field_parameters}.}
		\label{fig:np_strong_chirp_combined_field}
	\end{figure*}
	A similar overall behavior is observed for $\nu=5$ in Figs. \ref{fig:np_strong_chirp_strong_field}(d)--\ref{fig:np_strong_chirp_strong_field}(f), although the chip-induced modification are more pronounced. For $\lambda = 500\,m^{-1}$, increasing $b_1$ produces increasingly strong oscillations preceding the main positive-momentum peak. At $b_1 = 0.5\omega_1/\tau$, these structures become particularly prominent, while for $b_1 = \omega_1/\tau$ the oscillatory region extends over a broader momentum interval. For $\lambda = 10\,m^{-1}$, the chirp substantially redistributes the spectrum, producing pronounced oscillations around negative and near-zero momentum and modifying the dominant positive-momentum peak. The strongest spatial inhomogeneity, $\lambda = 2\,m^{-1}$, again gives a more localized distribution, but its peak position and amplitude are noticeably modified with increasing chirp. For $\nu=5$, the largest chirp also produces a more extended high-momentum tail, most clearly visible for $\lambda = 500\,m^{-1}$, indicating that chirping can redistribute the produced-particle yield towards larger final momenta.
	
	The corresponding dependence of the reduced total yield on $\lambda$ is shown in Fig. \ref{fig:N_strong_field}. For both values of $\nu$, the yield increases rapidly with increasing $\lambda$ and approaches saturation at sufficiently large $\lambda$. Thus, the strong-field yield is enhanced as the field becomes less spatially inhomogeneous. The effect of chirp is comparatively weak for $b_1 = 0.1\omega_1/\tau$, for which the yield remains close to the chirp-free result over the entire range of $\lambda$. Increasing the chirp to $b_1 = 0.5\omega_1/\tau$ produces a more visible enhancement, particularly for intermediate and large $\lambda$. The largest chirp, $b_1=\omega_1/\tau$, leads to the strongest enhancement of the total yield, with the difference becoming especially apparent as the yield approaches its saturation value. The dependence on $\nu$ is also evident from Fig. \ref{fig:N_strong_field}. For $\nu=5$, the curves corresponding to different chirp parameters exhibit considerably larger separations than those for $\nu=1$, along with an overall enhancement of the pair yield upon increasing $\nu$ from $1$ to $5$.
		
	\subsection{Two-color combined field $E(x,t)$}\label{subsec:strong_chirp_combined_field}

	\begin{figure*}[tbh]
		\centering
		\includegraphics[width=0.9\linewidth]{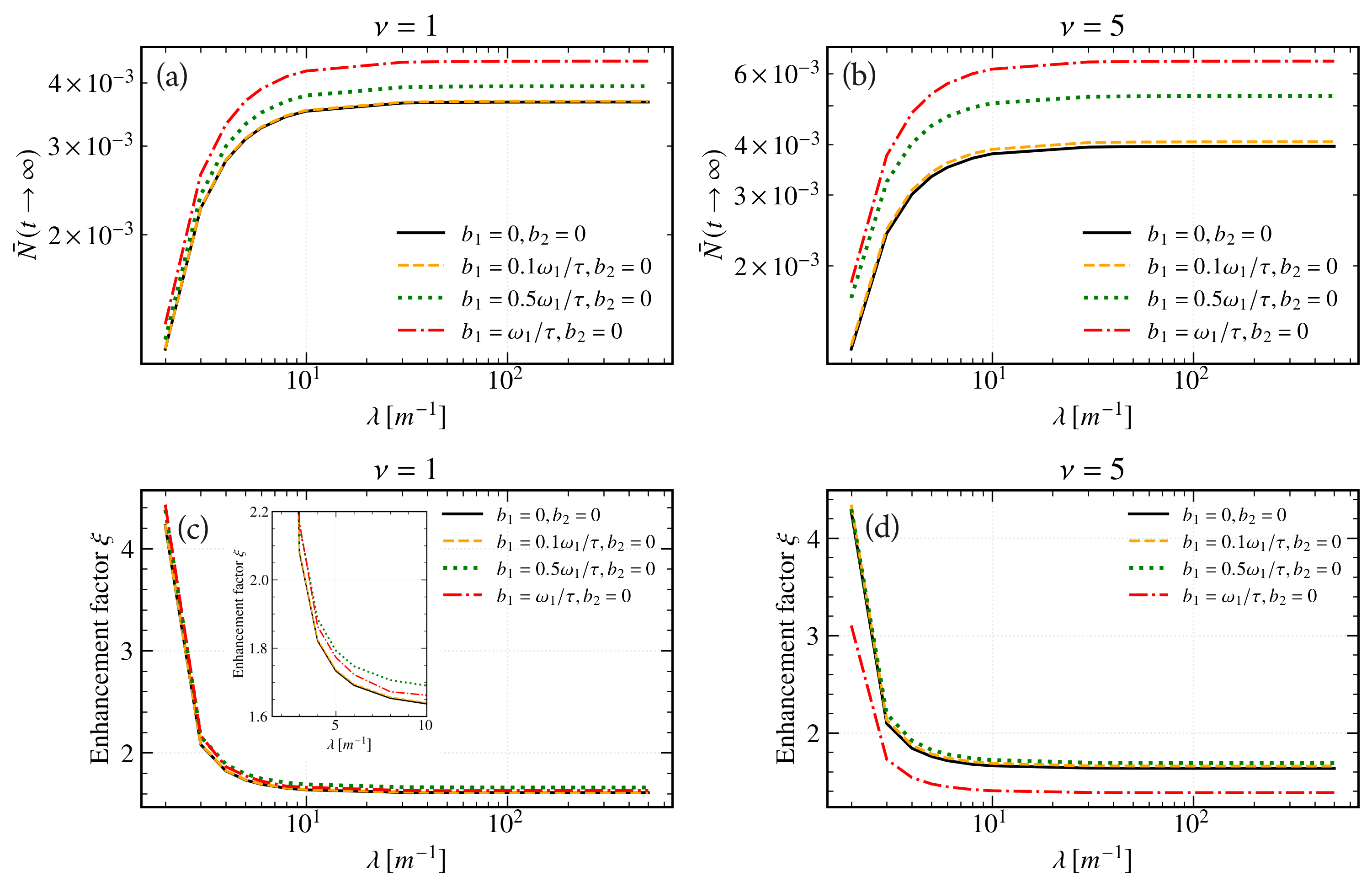}
		\caption{Reduced total yield $\bar{N}(t\rightarrow\infty)$ (panel (a) and (b)) and the corresponding enhancement factor (panel (c) and (d)) as a function of spatial scales for two-color combined field $E(x,t)$, with chirp applied to strong field component $E_{1s}(x,t)$ only. The inset in panel (c) shows a magnified view of the enhancement factor in the small-$\lambda$ region. The left and right columns correspond to $\nu=1$ and $\nu=5$, respectively. The other field parameters are given in Eq.~\eqref{eq:field_parameters}.}
		\label{fig:N_strong_chirp_combined_field}
	\end{figure*}
	In this subsection, we investigate the case of two-color combined field $E(x,t)$ with chirp applied to $E_{1s}(x,t)$ only, i.e., $b_1\neq0$ and $b_2=0$. The corresponding reduced momentum distribution and total yields are shown in Fig. \ref{fig:np_strong_chirp_combined_field}, and Figs. \ref{fig:N_strong_chirp_combined_field}(a) and \ref{fig:N_strong_chirp_combined_field}(b), respectively. These results allow us to examine how the chirp-induced modification of the strong field influences the dynamically assisted pair-production process.
	
	As shown in Fig. \ref{fig:np_strong_chirp_combined_field}, the momentum distribution retains a strong dependence on $\lambda$. Introducing chirp produces additional oscillatory structures, particularly on the low-momentum side of the main peak. These structures become increasingly pronounced as $b_1$ is increased and are especially evident for $\nu=5$. For the largest chirp, the momentum distribution develops multiple oscillations and additional structures extending towards higher momenta, indicating a stronger redistribution of the produced particles in momentum space. The effect is most pronounced for the larger spatial scales, whereas strong spatial inhomogeneity tends to confine the distribution to a narrower momentum region. Increasing $\nu$ from 1 to 5 further enhances these chirp-induced structures and produces more pronounced high-momentum tails. 
	
	The corresponding reduced total yields in Figs. \ref{fig:N_strong_chirp_combined_field}(a) and \ref{fig:N_strong_chirp_combined_field}(b) increase rapidly with $\lambda$ and approach saturation at larger spatial scales, consistent with the behavior observed for the chirp-free combined field case [Fig. \ref{fig:N_chirp_free_SWC_fields}(c)] and chirped strong-field case [Fig. \ref{fig:N_strong_field}]. At a fixed $\lambda$, increasing the chirp strength systematically enhances the total yield. This enhancement is more pronounced for $\nu=5$, for which the separation between the different chirp cases is considerably larger than for $\nu=1$. 
	\begin{figure*}[tbh]
		\centering
		\includegraphics[width=\textwidth]{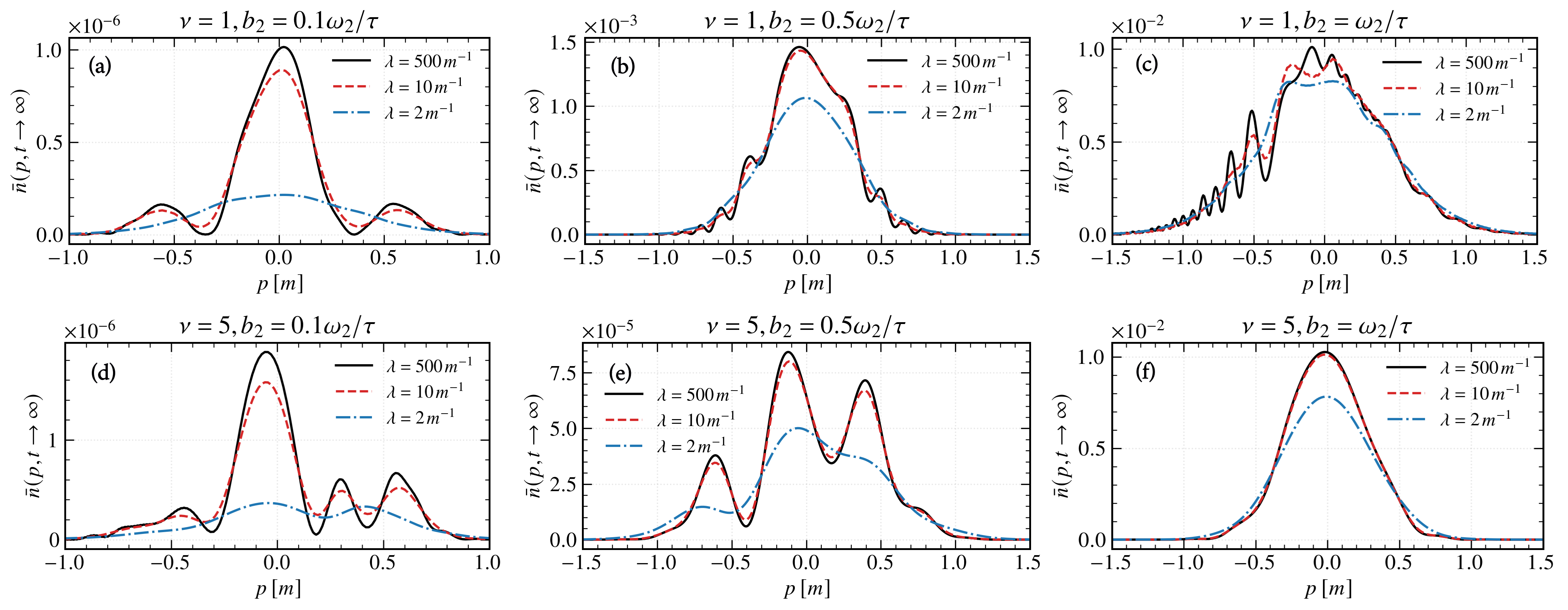}
		\caption{Reduced momentum distribution for different spatial scales in one-color weak field $E_{2w}(x,t)$. The first column [panels (a),(d)], second column [panels (b),(e)], and third column [panels (c),(f)] correspond to the chirp $b_2[\omega_2/\tau] = 0.1,0.5$ and $1$, respectively. The first row [panels (a), (b), and (c)] and second row [panels (d), (e), and (f)] represent the cases $\nu=1$ and $\nu=5$, respectively. The other field parameters are given in Eq.~\eqref{eq:field_parameters}.}
		\label{fig:np_weak_chirp_weak_field}
	\end{figure*}
	\begin{figure}[h]
		\centering
		\includegraphics[width=0.9\linewidth]{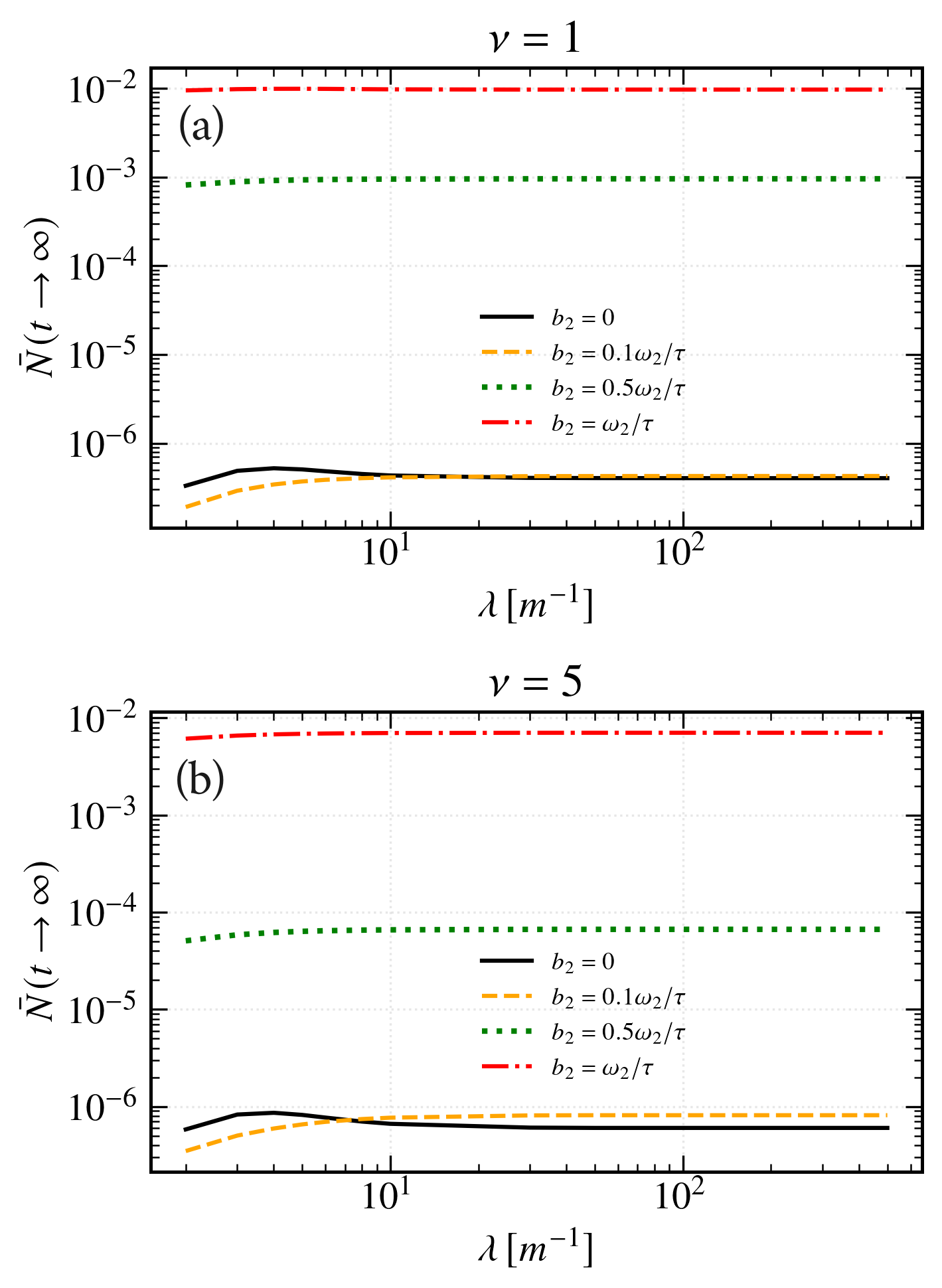}
		\caption{Reduced total yield for one-color weak field $E_{2w}(x,t)$ with different chirp as a function of spatial scales. Results are presented for $\nu=1$ (panel (a)) and $\nu=5$ (panel (b)). The other field parameters are given in Eq.~\eqref{eq:field_parameters}.}
		\label{fig:N_weak_field}
	\end{figure}
	To quantify the efficiency of the dynamical assistance, we consider the enhancement factor shown in Figs. \ref{fig:N_strong_chirp_combined_field}(c) and \ref{fig:N_strong_chirp_combined_field}(d). For all chirp values, the enhancement factor decreases rapidly with increasing $\lambda$, followed by a much slower variation and an approximately constant value in the weakly inhomogeneous regime, implying that the dynamically assisted enhancement is strongest for small spatial scales. An important feature is that increasing the chirp does not lead to a corresponding increase in the enhancement factor. Although the total combined-field yield increases with chirp, the enhancement factor generally decreases, particularly for $\nu=5$. This indicates that chirping enhances the strong-field pair production itself more efficiently than it enhances the additional contribution arising from the weak field. Consequently, the relative advantage of the dynamically assisted two-color configuration is reduced as the chirp becomes stronger.
	
	Overall, chirping the strong component provides control over both the magnitude and spectral structure of pair production. Increasing $b_1$ enhances the yield and produces more pronounced momentum-space structures, particularly for $\nu=5$. In contrast, the enhancement factor is largest for strongly spatially inhomogeneous fields and generally decreases with increasing $\lambda$ and chirp, highlighting the distinct roles of these two effects in determining the absolute yield and the relative strength of dynamical assistance.
	
	\section{Numerical results: Chirp applied to $E_{2w}(x,t)$ only}\label{sec:Numerical_results_weak_chirp}
	In this section, we present the results for the one-color weak field $E_{2w}(x,t)$ and the two-color combined field $E(x,t)$. In both the cases, we consider the chirping only for $E_{2w}(x,t)$, i.e., $b_1=0$ and $b_2 = 0.1\omega_2/\tau, 0.5\omega_2/\tau$ and $\omega_2/\tau$.
	\subsection{One-color weak field $E_{2w}(x,t)$} \label{subsec:weak_chirp_weak_field}
	
	In this subsection, we consider the results of the chirped one-color weak field $E_{2w}(x,t)$ with $\lambda$ and $\nu$ varied. Fig. \ref{fig:np_weak_chirp_weak_field} shows the corresponding reduced momentum distributions for three nonzero chirp parameters, $b_2 = 0.1\omega_2/\tau, 0.5\omega_2/\tau$ and $\omega_2/\tau$, while Fig. \ref{fig:N_weak_field} presents the corresponding reduced total yield as a function of the spatial scale. 
	
	The contrasting response to chirping in the strong- and weak-field regimes can be attributed to the different mechanisms governing pair production. As observed in previous sections, a small chirp, $b_1=0.1\omega_1/\tau$ [Fig. \ref{fig:np_strong_chirp_strong_field}(a),(d)], leaves the overall shape of the momentum distribution in the strong-field case largely unchanged from the unchirped case [Fig. \ref{fig:np_chirp_free_SWC_fields}(a),(d)]. In this regime, pair production is predominantly controlled by the field strength, while the dependence on the frequency enters mainly through the exponential suppression characteristic of subcritical fields. Consequently, a small modification of the instantaneous frequency due to chirping has only a minor effect on the momentum spectrum. In contrast, the weak-field regime is more sensitive to the frequency content of the external field, making the momentum distribution substantially more responsive even to relatively small chirps. For $b_2 = 0.1\omega_2/\tau$, the momentum distribution in Fig. \ref{fig:np_weak_chirp_weak_field} remain dominated by a central peak around $p\sim0$, with additional weaker structures appearing on either side. The distributions for $\lambda=500\,m^{-1}$ and $10\,m^{-1}$ are quite similar, whereas the strongly inhomogeneous case $\lambda=2\,m^{-1}$ exhibits a noticeable reduction in magnitude. Increasing the chirp to $b_2=0.5\omega_2/\tau$ substantially modifies the spectrum, producing multiple pronounced peaks and a broader momentum distribution, as seen in Figs. \ref{fig:np_weak_chirp_weak_field}(b) and \ref{fig:np_weak_chirp_weak_field}(e). For the largest chirp, $b_2 = \omega_2/\tau$, the spectrum becomes considerably broader and develops pronounced oscillatory structures, particularly for $\nu=1$. At the same time, the distributions corresponding to different spatial scales become increasingly similar in shape, including a reduced sensitivity to spatial inhomogeneity. Similar trends are observed for $\nu=5$, although the detailed peak structure and relative amplitudes are modified, demonstrating the additional influence of $\nu$ on the chirped weak-field dynamics. 
	 
	The corresponding total yield in Fig. \ref{fig:N_weak_field} exhibit a particularly strong dependence on the chirp parameter. For $b_2 = 0$ and $b_2=0.1\omega_2/\tau$, the yield remains of order $10^{-6}$ and shows a noticeable dependence on $\lambda$, especially at small $\lambda$. Increasing chirp to $b_2 = 0.5\omega_2/\tau$ enhances the yield by several orders of magnitude, while $b_2=\omega_2/\tau$ produces a further dramatic enhancement, reaching the $10^{-2}$ level. This strong increase reflects the increase in the instantaneous effective frequency induced by the chirp, which facilitates pair production through higher-energy multiphoton processes. The dependence on $\nu$ is comparatively modest for the smaller chirps, whereas at stronger chirping the overall yields and their detailed spectral structures are noticeably modified between $\nu=1$ and $\nu=5$. 
	
	An important feature of Fig. \ref{fig:N_weak_field} is the progressive reduction in the total yield's sensitivity to the spatial scale with increasing chirp. This behavior contrasts with the strong field case considered previously (cf. Fig. \ref{fig:N_strong_field}), where the reduced total yield remains sensitive to variations in the spatial scale, irrespective of the value of chirp. For weak chirping, decreasing $\lambda$ reduces the available field region and hence the field energy contribution to pair production, leading to a stronger suppression of the yield for spatially localized fields. This behavior is consistent in the current case where reduced total yield remains significantly influenced by the spatial extent of the weak fields when chirp is small i.e., $b_2 = 0.1\omega_2/\tau$. As the chirp becomes stronger, however, the increasing effective frequency allows pair production to proceed predominantly through higher-energy photon absorption. Such a mechanism depends less strongly on the total spatial extent and field energy of the localized region, thereby making the yield progressively less sensitive to $\lambda$. In the strongly chirped case, the curves therefore become nearly flat over the considered spatial scales. A similar trend is also observed for $\nu=5$. In addition, a slight enhancement in the total yield is observed as $\nu$ increases from $1$ to $5$, with the enhancement being most pronounced, reaching an order of unity, at the intermediate chirp parameter, $b_1=0.5\omega_2/\tau$.
	
	Overall, chirping the weak field has a much stronger impact on the absolute production yield than the spatial-scale dependence. Increasing $b_2$ enhances the yield dramatically and transforms the momentum spectrum from a predominantly central peak into a broad, multi-peak and oscillatory structure. At the same time, the increasing effective frequency progressively weakens the dependence on spatial inhomogeneity. These results are important for the subsequent two-color analysis, where the chirped weak field provides the high-frequency component responsible for dynamical assistance.
	\subsection{Two-color combined field $E(x,t)$}\label{subsec:weak_chirp_combined_field}
\begin{figure*}[tbh]
		\centering
		\includegraphics[width=\textwidth]{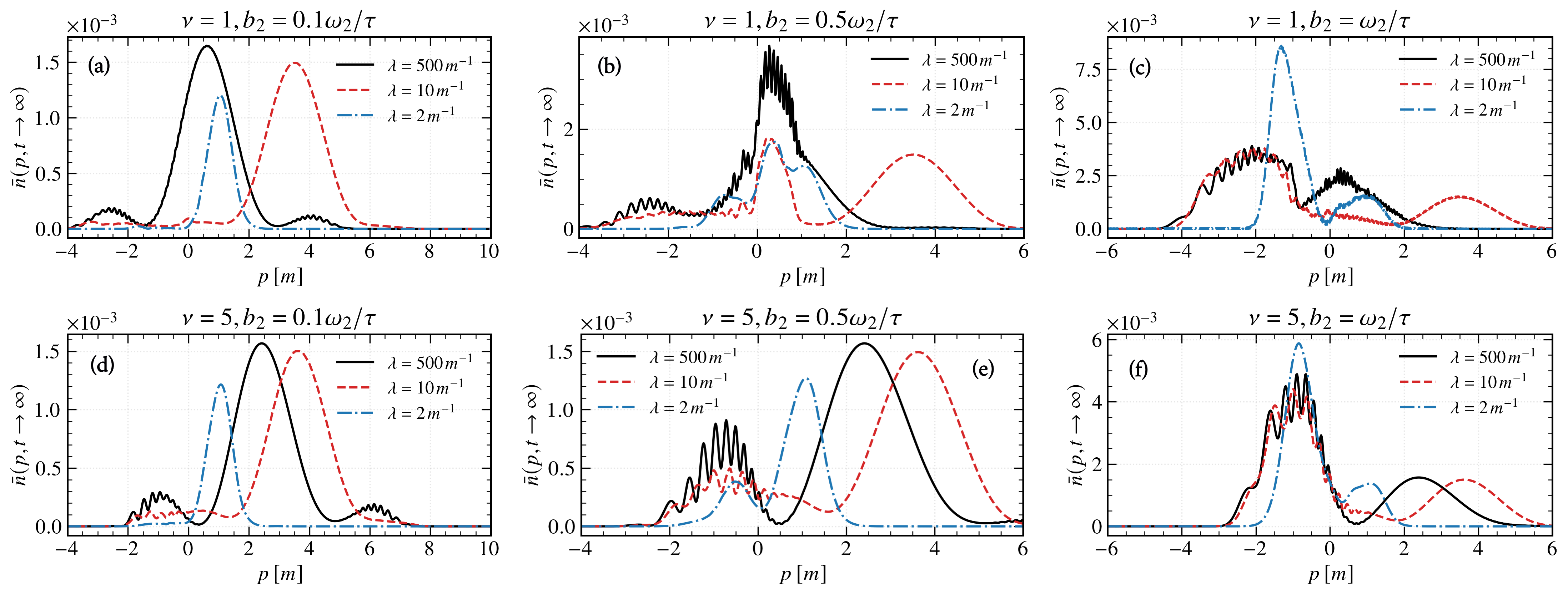}
		\caption{Reduced momentum distribution for different spatial scales in two-color combined field $E(x,t)$ with chirp applied to $E_{2w}(x,t)$ only. The first column [panels (a),(d)], second column [panels (b),(e)], and third column [panels (c),(f)] correspond to the chirp $b_2[\omega_2/\tau] = 0.1,0.5$ and $1$, respectively. The first row [panels (a), (b), and (c)] and second row [panels (d), (e), and (f)] represent the cases $\nu=1$ and $\nu=5$, respectively. The other field parameters are given in Eq.~\eqref{eq:field_parameters}.}
		\label{fig:np_weak_chirp_combined_field}
	\end{figure*}

\begin{figure*}[tbh]
		\centering
		\includegraphics[width=0.9\linewidth]{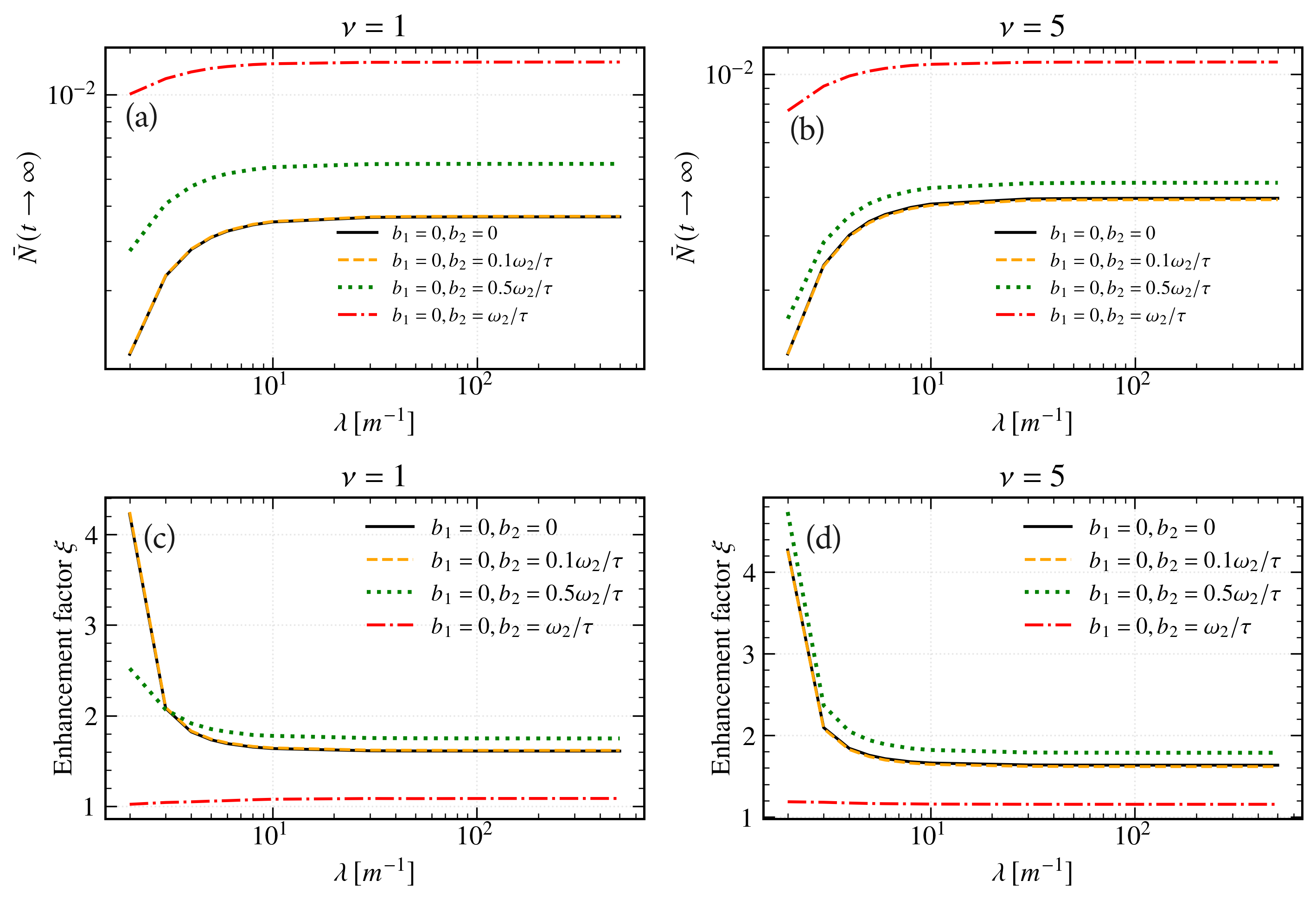}
		\caption{Reduced total yield $\bar{N}(t\rightarrow\infty)$ (panel (a) and (b)) and the corresponding enhancement factor (panel (c) and (d)) as a function of spatial scales for two-color combined field $E(x,t)$, with chirp applied to weak field component $E_{2w}(x,t)$ only. The left and right columns correspond to $\nu=1$ and $\nu=5$, respectively. The other field parameters are given in Eq.~\eqref{eq:field_parameters}.}
		\label{fig:N_weak_chirp_combined_field}
	\end{figure*}

	We now consider the two-color combined field $E(x,t) = E_{1s}(x,t) + E_{2w}(x,t)$, with chirp applied only to the weak field i.e., $b_1 = 0$ and $b_2 = 0.1\omega_2/\tau, 0.5\omega_2/\tau,$ and $\omega_2/\tau$. The corresponding reduced momentum distributions and total yields, together with the enhancement factor, are shown in Figs. \ref{fig:np_weak_chirp_combined_field} and \ref{fig:N_weak_chirp_combined_field}, respectively. 
	
	For the smallest chirp, $b_2=0.1\omega_2/\tau$, the momentum distributions in Fig. \ref{fig:np_weak_chirp_combined_field} remain qualitatively similar to the chip-free two-color case [Fig. \ref{fig:np_chirp_free_SWC_fields}(c),(f)], with a dominant peak at positive momentum. As $b_2$ is increased to $0.5\omega_2/\tau$, pronounced oscillatory structure develop, particularly for $\lambda=500\,m^{-1}$, while the dominant positive-momentum peak is shifted and broadened. For the largest chirp, $b_2=\omega_2/\tau$, the momentum distributions are strongly modified and exhibit several oscillatory and secondary structures. The dependence on $\lambda$ is also reduced, with the distributions for different spatial scales becoming increasingly similar in the strongly chirped regime. This behavior is consistent with the enhanced effective frequency of the weak field, which makes the production process less dependent on the spatial extent of the field. The changes are qualitatively similar for both $\nu=1$ and $\nu=5$, although the detailed peak positions and oscillatory structures differ.
	
	The reduced total yield in Figs. \ref{fig:N_weak_chirp_combined_field}(a) and \ref{fig:N_weak_chirp_combined_field}(b) increases rapidly with increasing $\lambda$ and approaches saturation in the weakly inhomogeneous regime. More importantly, the yield is strongly enhanced by increasing $b_2$. The enhancement is relatively moderate for $b_2=0.1\omega_2/\tau$, becomes much stronger for $b_2=0.5\omega_2/\tau$, and reaches its largest value for $b_2=\omega_2/\tau$, where the yield increases by more than an order of magnitude compared with the weakly chirped cases. This strong sensitivity to $b_2$ reflects the role of the chirped high-frequency field in providing additional energy for pair creation. The dependence on $\nu$ is comparatively weaker: the overall trends are similar for $\nu=1$ and $\nu=5$, although the yields for $\nu=1$ are slightly larger for the stronger chirps.
	
	An interesting feature is the decreasing sensitivity of the total yield to $\lambda$ as $b_2$ increases similar to the one-color weak field discussed in Subsec. \ref{subsec:weak_chirp_weak_field}. For small chirp, spatial inhomogeneity produces a noticeable suppression of the yield at small $\lambda$. With increasing chirp, however, the curves become progressively flatter. This is due to the increasing effective frequency of the weak field, which shifts pair production toward higher-energy channels and reduces its sensitivity to spatial localization, thereby partially compensating for the suppression due to spatial inhomogeneity.
	
	The enhancement factor in Figs. \ref{fig:N_weak_chirp_combined_field}(c) and \ref{fig:N_weak_chirp_combined_field}(d) provides an important complementary observation. For $b_2=0.1\omega_2/\tau$, the enhancement factor is largest at small spatial scales and decreases rapidly with increasing $\lambda$, eventually approaching an approximately constant value. A similar spatial dependence is retained for $b_2=0.5\omega_2/\tau$, although the magnitude of the enhancement is reduced. For the largest chirp, $b_2=\omega_2/\tau$, the enhancement factor is close to unity and shows only a weak dependence on $\lambda$. Thus, although chirping the weak field dramatically increases the absolute pair-production yield, it simultaneously reduces the relative dynamical-assistance enhancement. This occurs because the chirped weak field itself becomes highly efficient at producing pairs, so the additional gain obtained by combining it with the strong field becomes comparatively smaller.
	
	Overall, chirping the weak field produces a strong enhancement of the pair-production yield and substantially modifies the momentum spectrum, while progressively weakening its sensitivity to spatial inhomogeneity. The effect is particularly pronounced for the largest $b_2$ whereas the dependence on $\nu$ remains comparatively moderate. The reduction of the enhancement factor with increasing chirp further indicates that the dominant role of a strongly chirped weak field is to enhance direct pair production through its increased effective frequency, rather than to further amplify the relative dynamical assistance from the two-color configuration.
	
	\section{Numerical results: Chirp applied to both $E_{1s}(x,t)$ and $E_{2w}(x,t)$}\label{sec:Numerical_results_both_chirp}
\begin{figure*}[tbh]
		\centering
		\includegraphics[width=\textwidth]{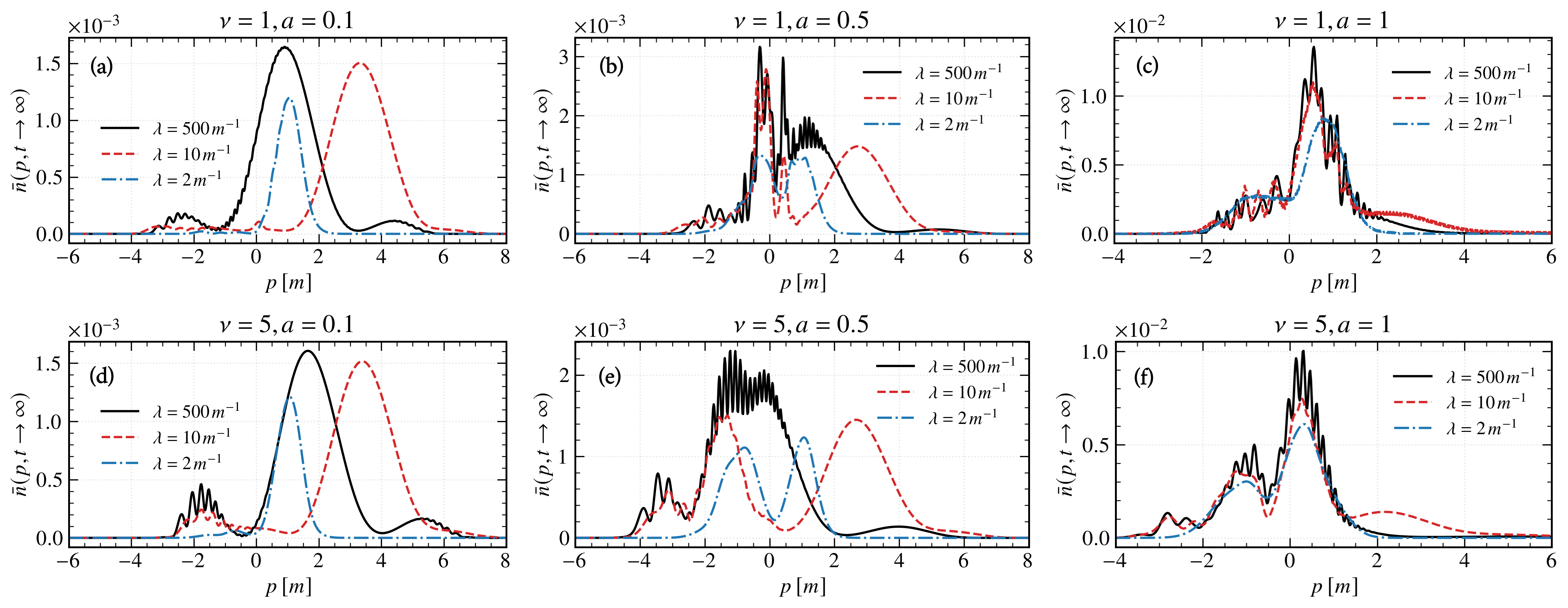}
		\caption{Reduced momentum distribution for different spatial scales in two-color combined field $E(x,t)$ with chirp applied to both $E_{1s}(x,t)$ and $E_{2w}(x,t)$. The first column [panels (a),(d)], second column [panels (b),(e)], and third column [panels (c),(f)] correspond to the chirp control parameter $a = 0.1,0.5$ and $1$, respectively, which define the chirp parameters $b_1$ and $b_2$ as $b_i = a\omega_i/\tau$ where $( i = 1,2)$. The first row [panels (a), (b), and (c)] and second row [panels (d), (e), and (f)] represent the cases $\nu=1$ and $\nu=5$, respectively. The other field parameters are given in Eq.~\eqref{eq:field_parameters}.}
		\label{fig:np_both_chirps_combined_field}
	\end{figure*}

\begin{figure*}[tbh]
		\centering
		\includegraphics[width=0.9\linewidth]{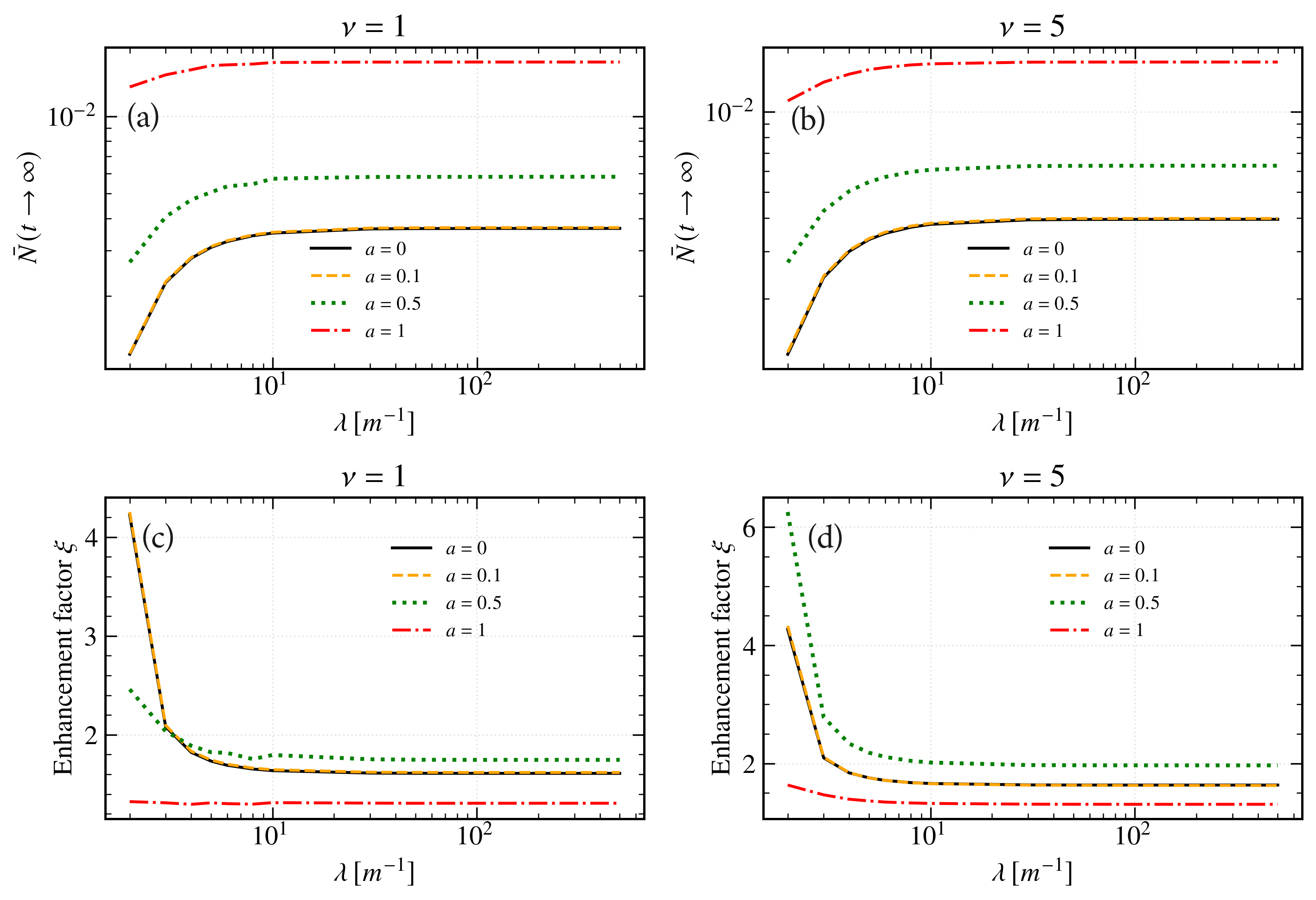}
		\caption{Reduced total yield $\bar{N}(t\rightarrow\infty)$ (panel (a) and (b)) and the corresponding enhancement factor (panel (c) and (d)) as a function of spatial scales for two-color combined field $E(x,t)$, with chirp applied to both strong and weak field components. The chirp is applied via the chirp control parameter $a = 0,0.1,0.5$ and $1$, which define the chirp parameters $b_1$ and $b_2$ as $b_i = a\omega_i/\tau$ where $( i = 1,2)$. The left and right columns correspond to $\nu=1$ and $\nu=5$, respectively. The other field parameters are given in Eq.~\eqref{eq:field_parameters}.}
		\label{fig:N_both_chirps_combined_field}
	\end{figure*}

	In this section, we consider the two-color combined field $E(x,t) = E_{1s}(x,t) + E_{2w}(x,t)$, with chirping applied simultaneously to both field components. In this case, the chirp parameter are chosen according to $b_i = a\omega_i/\tau, i=1,2$, with $a=0.1,0.5$, and $1$. The corresponding reduced momentum distributions and reduced total yields are shown in Figs. \ref{fig:np_both_chirps_combined_field} and \ref{fig:N_both_chirps_combined_field}, respectively. 
	
	For a relatively small chirp, $a=0.1$, the momentum distribution in Figs. \ref{fig:np_both_chirps_combined_field}(a) and \ref{fig:np_both_chirps_combined_field}(d) remain qualitatively similar to those of the chrip-free combined field [Fig. \ref{fig:np_chirp_free_SWC_fields}(c),(f)], although their peak positions and amplitudes are slightly modified. As $a$ is increased to $0.5$, pronounced oscillatory structure develop, particularly on the low-momentum side of the dominant positive-momentum peak. These structures become even more pronounced for $\nu=5$, indicating a stronger sensitivity of the momentum spectrum to the simultaneous frequency variation of the two components. For $a=1$, the spectrum is strongly reshaped and develops several closely spaced structures around the central momentum region, together with broader momentum tails. The distributions corresponding to different spatial scales also become increasingly similar as the chirp is increased, showing that strong chirping reduces the sensitivity of the momentum spectrum to spatial inhomogeneity. This behavior is consistent with the results obtained when chirp is applied to the weak field alone, where the increased effective frequency reduces the relative importance of the spatial extent of the field.
	
	The reduced total yields in Figs. \ref{fig:N_both_chirps_combined_field}(a) and \ref{fig:N_both_chirps_combined_field}(b) further demonstrate the strong effect of simultaneous chirping. For both $\nu=1$ and $\nu=5$, the yield increases rapidly with increasing $\lambda$ and approaches saturation in the weakly inhomogeneous regime. Increasing $a$ produces a systematic and substantial enhancement of the total yield, with the largest values obtained for $a=1$. Thus, simultaneous chirping of the strong and weak components is particularly effective in increasing the absolute pair-production yield. The enhancement in the total yield from $\nu=1$ to $\nu=5$ is barely noticeable over most of the parameter range. At the largest chirp, $a=1$, however, the yield enhancement is modestly greater for $\nu=1$ than for $\nu=5$.

	The enhancement factor shown in Figs. \ref{fig:N_both_chirps_combined_field}(c) and \ref{fig:N_both_chirps_combined_field}(d) reveals a different aspect. For $\nu=1$, the enhancement factor is largest for $a=0$ and $a=0.1$ at small spatial scales and decreases rapidly with increasing $\lambda$, eventually approaching an approximately constant value. In contrast, for $\nu=5$, the enhancement factor is largest at the intermediate chirp, $a=0.5$. For the largest chirp, $a=1$, the enhancement factor is close to unity and exhibits only a weak dependence on the spatial scale for both $\nu=1$ and $\nu=5$. Hence, as observed  in Figs. \ref{fig:N_weak_chirp_combined_field} and \ref{fig:N_both_chirps_combined_field}, stronger chirping can greatly increase the absolute pair-production yield while reducing the relative enhancement associated with two-color dynamical assistance. This is because the chirped fields themselves become increasingly efficient at producing pairs, making the additional contribution from combining the two components relatively less significant.
	
	Comparing the three chirping configurations considered above, a consistent picture emerges. Chirping the strong field primarily modifies the momentum distribution and enhances the strong-field yield, whereas chirping the weak field has a particularly strong impact on the absolute production rate through its rapidly increasing effective frequency. When both components are chirped simultaneously, these effects act together, resulting in the largest overall yields and substantial modifications of the momentum spectrum. At the same time, the enhancement factor demonstrates that a larger absolute yield does not necessarily correspond to a larger dynamical-assistance enhancement.
	
	These results also suggest a possible route toward optimal control of pair production. Rather than maximizing the total yield alone, one could systematically optimize the chirp strengths $b_1,b_2$, and the spatial scale $\lambda$ using the enhancement factor as a figure of merit. Such an analysis could identify parameter regions where the absolute yield and the relative dynamical-assistance enhancement are simultaneously optimized, providing a natural direction for further investigation.
	
	\section{Relative phase effects}\label{sec:non-zero_relative_phase}
	\begin{figure*}[tbh]
		\centering
		\includegraphics[width=\linewidth]{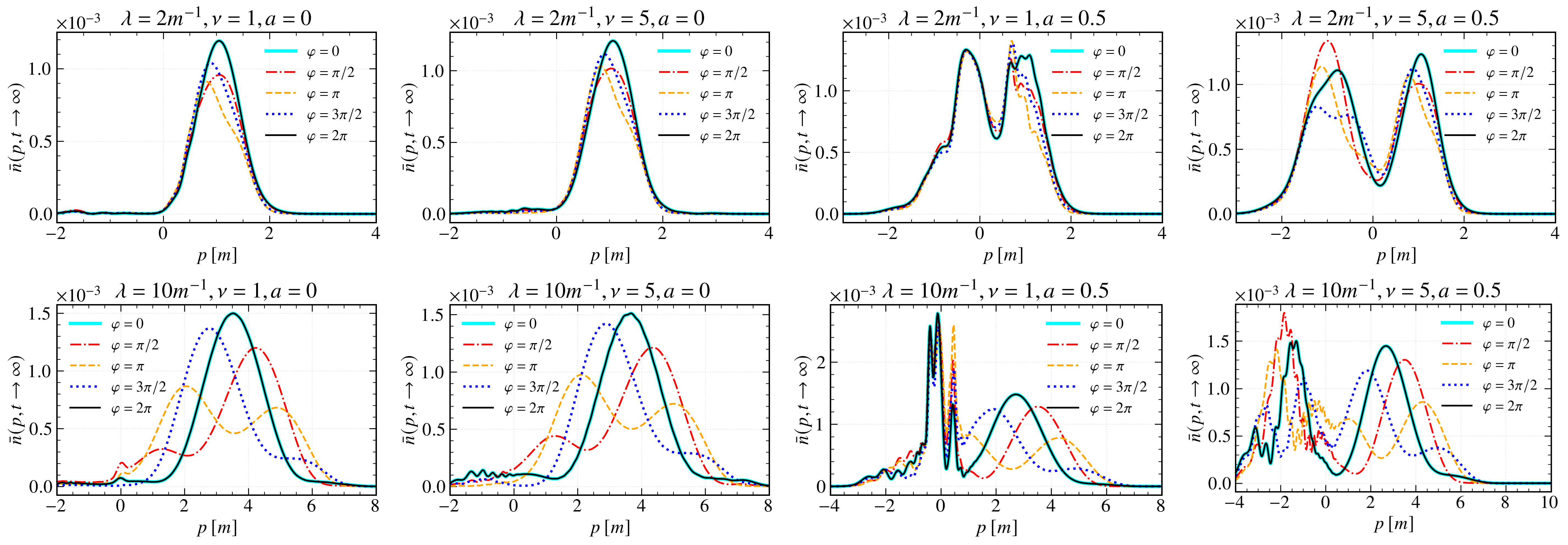}
		\caption{Reduced momentum distribution for the two-color combined field $E(x,t)=E_{1s}(x,t)+E_{2w}(x,t)$ for different relative phases $\varphi$, spatial scales, and values of $\nu$. The first and second rows correspond to $\lambda=2~m^{-1}$ and $\lambda=10~m^{-1}$, respectively. The first and second columns show the results for the chirp-free case, $a=0$, with $\nu=1$ and $\nu=5$, respectively, while the third and fourth columns correspond to $a=0.5$ for $\nu=1$ and $\nu=5$, respectively. The chirp parameters are defined as $b_i=a\omega_i/\tau$ $(i=1,2)$. The remaining field parameters are given in Eq. \eqref{eq:field_parameters}.}
		\label{fig:np_combined_relative_phase}
	\end{figure*}
	In the preceding sections, we have restricted the relative phase between the strong and weak field components to zero. We now extend the analysis by introducing a finite relative phase $\varphi$ between the two color components and investigate its influence on the pair production. Fig. \ref{fig:np_combined_relative_phase} presents the reduced momentum distribution for $\varphi = 0, \pi/2, \pi, 3\pi/2 $, and $2\pi$, considering two representative spatial scales, $\lambda=2\,m^{-1}$ and $10\,m^{-1}$, and both the chirp-free case $a = 0$ and the case of simultaneous chirping $a = 0.5$, with $b_i = a\omega_i/\tau$. The results are shown for $\nu=1$ and $\nu=5$.
	
	In the chirp-free case, the momentum distribution exhibits a clear dependence on the relative phase, although the strength of this dependence varies more with the spatial scale $\lambda$ and less with super-Gaussian order $\nu$. For $\lambda=2\,m^{-1}$, the spectra for different phases remain relatively similar in shape, with the main peak occurring around $p\sim0$; the primary phase dependence is reflected in its magnitude and position. For the larger spatial scale, $\lambda=10\,m^{-1}$, the phase dependence becomes considerably more pronounced, with noticeable shifts in the position of the dominant peak and changes in its amplitude and width. This indicates that the relative phase can substantially modify the momentum-space redistribution of the produced pairs when the field is less spatially localized. The results for $\nu=5$ show qualitatively similar behavior, although the detailed peak positions and spectral widths are negligibly modified.
	
	When chirping is applied simultaneously to both field components, the phase dependence becomes substantially stronger. For $a = 0.5$, the relatively smooth distributions of the chirp-free case develop pronounced oscillatory and multi-peak structures. This behavior is especially evident for $\lambda=10\,m^{-1}$, where different values of $\varphi$ lead to significant changes in the relative amplitudes and positions of the multiple spectral peaks. Even for the strongly inhomogeneous case $\lambda=2\,m^{-1}$, where the chirp-free spectra are comparatively similar, chirping produces clear phase-dependent oscillations. The enhanced sensitivity can be attributed to the additional interference introduced by the chirped temporal phases of the two field components, which modify the relative phase accumulated during the pair-production process. Thus, chirping not only reshapes the momentum spectrum but also make the spectrum more sensitive to the relative phase. 
	
	Increasing $\nu$ from 1 to 5 modifies the peak locations, widths, and fine structures of the spectra for the same $\lambda$ and $\varphi$. In the chirped case, this difference becomes more pronounced, with $\nu=5$ exhibiting stronger fine structure and a broader redistribution of the momentum spectrum. Nevertheless, the qualitative phase dependence remains the same, indicating that the relative phase acts as an additional control parameter independently of the overall field configuration. 
	
	The spectra at $\varphi=0$ and $2\pi$ essentially coincide in all the considered cases, as expected from the periodicity of the relative phase. More generally, the variation observed as $\varphi$ is changed demonstrate that the two-color pair-production process is sensitive not only to the field amplitudes, frequencies, spatial scale, and chirp, but also to the phase relation between the two components. In particular, the pronounced changes in the chirped case suggest that the relative phase can be used to control the detailed momentum-space structure of the produced pairs. This provides an additional degree of freedom for controlling and potentially optimizing pair production in spatially inhomogeneous two-color fields.
	
	\section{Conclusions and Discussions}\label{sec:conclusion}
	In this work, we have investigated dynamically assisted electron-positron pair production in spatially inhomogeneous electric fields with super-Gaussian temporal profiles, focusing on the combined effects of spatial localization, pulse shape, frequency chirping, and the relative phase between the strong and weak field components.  Using the (1+1)-dimensional Dirac-Heisenberg-Wigner formalism, we have analyzed both the reduced momentum distributions and the reduced total particle yields for three field configurations: a one-color strong-slowly varying field, a one-color weak-rapidly varying field, and their dynamically assisted two-color combinational field. In particular, we have systematically compared chirping of strong field, chirping of the weak field, and simultaneous chirping of both components, thereby providing a unified picture of how these control parameters modify dynamically assisted pair production. The main findings of this study are summarized as follows:
	\begin{enumerate}[label=(\roman*)]
		\item For the chirp-free configurations, spatial inhomogeneity strongly influences both the reduced momentum distribution and reduced total yield. The strong-field spectrum becomes narrower with decreasing spatial scale, while the weak-field distribution evolves from a two-lobed structure toward a broader, suppressed spectrum. The combined field exhibits a distinct momentum distribution, characterized by a pronounced positive-momentum peak and oscillatory structures, demonstrating the dynamical assistance arising from the interplay of the two components. Increasing the super-Gaussian order from $\nu=1$ to $\nu=5$ mainly enhances the spectral structures, with a relatively modest effect on the total yield. The total yield increases with spatial scale and saturates in the weakly inhomogeneous regime, while the enhancement factor is largest for strongly inhomogeneous fields and decreases with increasing spatial scale $\lambda$, approaching an approximately constant value for large $\lambda$. A slightly larger enhancement is obtained when flat-top character of the temporal envelope is increased.
		\item Frequency chirping provides an additional and highly effective mechanism for modifying the pair-production dynamics. When chirping is applied to the strong field alone, the total yield increases with the chirp strength, while the momentum distribution develops increasingly pronounced oscillatory structures. For the higher-order super-Gaussian pulse, the strongest chirp additionally produces an extended high-momentum tail. In the combined field, strong-field chirping similarly enhances the absolute yield and generates increasingly complex momentum-space structures. However, the enhancement factor does not increase correspondingly; rather, the strongest chirp enhancement factor curve $b_1=\omega_1/\tau$ stays lowest among all the considered chirp parameters.
		\item A considerably stronger response is obtained when the chirp is applied to the weak, high-frequency component. Even relatively small chirping substantially modifies the weak-field momentum spectrum, while stronger chirping produces broad, multi-peak, and oscillatory distributions. The total yield can increase by several orders of magnitude as the chirp is increased, reflecting the increasing effective frequency and the resulting access to higher-energy multiphoton production channels. At the same time, the dependence of the yield on the spatial scale becomes progressively weaker with increasing chirp. Thus, sufficiently strong frequency chirping can partially compensate for the suppression associated with spatial localization by shifting the dominant production mechanism towards higher-frequency channels. The same behavior is reflected in the dynamically assisted two-color configuration. Chirping the weak component produces a substantial increase in the absolute pair-production yield and strongly reshapes the momentum distribution, but the corresponding enhancement factor decreases as the chirp becomes stronger. For the largest weak-field chirp considered here, the enhancement factor approaches unity and becomes only weakly dependent on the spatial scale. This indicates that a strongly chirped weak field becomes sufficiently efficient at producing pairs on its own that the additional gain obtained by combining it with the strong field becomes comparatively less significant. Hence, a large absolute yield does not necessarily imply a large dynamical-assistance enhancement.
		\item When both field components are chirped simultaneously, the effects of the two individual chirps act together. This configuration produces the largest absolute pair-production yields among the chirping scenarios considered in this work and leads to substantial modifications of the momentum spectra, including closely spaced structures and broader momentum tails at stronger chirping. The enhancement of the total yield becomes particularly pronounced for the higher-order super-Gaussian pulse. Nevertheless, the enhancement factor again exhibits a contrasting behavior: it decreases with increasing chirp and becomes close to unity for the largest chirp considered. Thus, simultaneous chirping is highly effective when the objective is to maximize the absolute number of created pairs, but it does not necessarily maximize the relative benefit of dynamical assistance. Thus, our study of the enhancement factor reveals two different optimization objectives. This distinction is particularly important when identifying favorable field configurations for controlled pair production.
		\item Our study finds that the enhancement factor generally decreases with increasing field width, indicating that the enhancement of pair production becomes weaker as the field becomes more spatially extended. However, Li \textit{et al.} demonstrated that this trend does not persist indefinitely: for sufficiently narrow fields, the enhancement factor can also decrease with decreasing field width, resulting in one or two maxima near the characteristic field-width thresholds associated with the strong- and weak-field regimes \cite{li2021enhanced}. Importantly, their results, obtained using a $\cos^4(t/\tau)$ temporal envelope rather that the super-Gaussian temporal profile considered here, further demonstrate that the detailed behavior of the enhancement factor depends on the temporal structure of the external field under consideration. In addition, the magnitude and location of the enhancement are sensitive to the specific field parameters. These observations indicate that the pair-production yield cannot be optimized simply by increasing or decreasing the spatial width of the field. Rather, an appropriate combination of spatial, temporal, and field-strength parameters may be required to maximize the production rate. This motivates further systematic studies aimed at optimizing the relevant field parameters and, ultimately, identifying an optimal configuration for pair production in space-time-dependent electric fields.
		\item Finally, we have demonstrated that the relative phase between the strong and weak components provides an additional degree of control. Changing the phase modifies the momentum-space structure of the produced pairs, with the effect becoming more pronounced for larger spatial scales and in the presence of chirping. The stronger phase sensitivity in the chirped case can be associated with the additional interference introduced by the chirped temporal phases, which modify the relative phase accumulated during the pair-production process. Although the qualitative phase dependence remains robust for different super-Gaussian orders, increasing flat-top character enhances the associated fine structures and momentum redistribution. 	
	\end{enumerate}
	Taken together, our results demonstrate that pulse shape, spatial localization, frequency chirping, and relative phase provide complementary means of controlling dynamically assisted pair production. Importantly, maximizing the absolute yield does not necessarily maximize the dynamical-assistance enhancement: strong chirping can substantially increase the yield while reducing the enhancement factor, whereas strong spatial inhomogeneity generally favors larger enhancement. Thus, systematic optimization over temporal, spatial, spectral, and phase parameters is required to identify favorable configurations. Our results provide a useful reference for optimal control of dynamically assisted pair production in space-time-dependent electric fields.
	
	\section*{Acknowledgments}
	AJ acknowledges fruitful discussions with Dr. M. Ababekri. The authors are grateful to Dr. C. Kohlf\"urst for making the numerical implementation available through his PhD thesis \cite{KohlfurstPhD}, which was helpful in developing the computational framework used in this work.
	
	\bibliography{ref} 
\end{document}